\documentclass[manuscript,nonacm,screen,natbib=false]{acmart}
\usepackage{courier}
\usepackage[inline]{enumitem}
\usepackage{algorithm}
\usepackage[noend]{algpseudocode}
\usepackage{xspace}
\usepackage{hyperref}
\usepackage{cleveref}
\usepackage{multirow}
\usepackage[flushleft]{threeparttable}
\usepackage{adjustbox}
\usepackage{subcaption}
\usepackage{tikz}
\usetikzlibrary{cd, shapes, arrows, positioning, intersections}
\tikzset{
  universe/.style={
    diamond,
    draw,
    shape aspect=1,
    anchor=west, 
  },
  upwardclosed/.style={
    isosceles triangle,
    draw,
    anchor=right corner, 
    shape border rotate=-90,
  },
  convex/.style={
    diamond,
    draw,
    shape aspect=.6,
    anchor=west, 
  },
  permit/.style={
    fill=blue!20,
    opacity=0.8,
    draw=black,
  },
  deny/.style={
    fill=white,
    opacity=0.8,
    draw=black,
  },
  hasse/.style={
    blue, 
    ultra thick, 
    densely dashed,
  },
}
\newcommand{\hasseFont}[1]{\textcolor{blue}{\mathbf{\ensuremath{#1}}}}

\newtheorem{definition}{Definition}
\newtheorem{lemma}{Lemma}
\newtheorem{theorem}{Theorem}
\newtheorem{example}{Example}

\newlist{thmenum}{enumerate}{1}
\setlist[thmenum]{label=(\roman{thmenumi}), ref=\thetheorem(\roman{thmenumi}),noitemsep}
\Crefname{thmenumi}{Theorem}{Theorems}

\newcommand{\tuple}[1]{\ensuremath{\langle#1\rangle}}
\newcommand{\set}[1]{\ensuremath{\{#1\}}}
\newcommand{\id}[1]{\texttt{#1}}
\newcommand{\deny}{\id{DENY}\xspace}
\newcommand{\permit}{\id{PERMIT}\xspace}
\newcommand{\complementedexp}{\ensuremath{\alpha^-}\xspace}
\newcommand{\uncomplementedexp}{\ensuremath{\alpha^+}\xspace}
\newcommand{\compuncompexp}{\ensuremath{\alpha^\pm}\xspace}
\newcommand{\allconditions}{\ensuremath{\mathcal{C}}\xspace}
\newcommand{\allexpressions}{\ensuremath{\Phi}\xspace}
\newcommand{\allminterms}[1][\mathcal{C}]{\ensuremath{\mathcal{M}_{#1}}\xspace}
\newcommand{\upwardclosure}[1]{#1^{\uparrow}}
\newcommand{\downwardclosure}[1]{#1^{\downarrow}}
\newlist{policy}{itemize}{1}
\setlist[policy]{label=,itemjoin={{,}},afterlabel=\unskip{{~}}}
\newlist{enumcase*}{enumerate*}{1}
\setlist[enumcase*]{
  label=(\thesubsection-\arabic*),
  before=\unskip{: },
  itemjoin={{; }},
  itemjoin*={{, or }}
}
\crefname{enumcase*i}{case}{cases}

\AtBeginDocument{%
  }

\RequirePackage[
  datamodel=acmdatamodel,
  maxbibnames=99,
  ]{biblatex}

\begin{document}

\title{Expressiveness and Convertibility of Rule-Based Access Control Policies}

\author{Amir Masoumzadeh}
\orcid{0000-0003-0647-7785}
\email{amasoumzadeh@albany.edu}
\author{Paliath Narendran}
\orcid{0000-0003-4521-5892}
\email{pnarendran@albany.edu}
\affiliation{%
  \institution{University at Albany, SUNY}
  \city{Albany}
  \state{New York}
  \country{USA}
}

\author{Padmavathi Iyer}
\orcid{0000-0001-9611-451X}
\email{riyer@drury.edu}
\affiliation{%
  \institution{Drury University}
  \city{Springfield}
  \state{Missouri}
  \country{USA}
}


\begin{abstract}
Rule-based access control models such as attribute-based access control support flexible policies based on rules with conditional expressions.
The design choices in those models such as what is allowed in the conditional expressions or supporting deny rules may impact the range of the policies that are expressible.
In this paper, we propose a systematic approach to investigate the semantics and compare the expressiveness of various rule-based access control models that can be built based on different design choices of rule structure and rule composition.
Furthermore, we investigate the problem of convertibility of a policy from one rule-based model to another. 
This question is critical for policy migration to ensure that a given policy instance in a source model is expressible in a target model.
Some of our major results in this paper include well-formedness properties for policy semantics, expressiveness hierarchy of models, and computational complexity of policy convertibility, all having important implications on the design and application of rule-based access control policy models.
\end{abstract}

\ccsdesc[500]{Security and privacy~Formal security models}
\ccsdesc[500]{Security and privacy~Authorization}
\ccsdesc[500]{Security and privacy~Access control}

\keywords{access control policy model, rule-based access control, expressiveness, policy conversion}


\maketitle

\section{Introduction}
\label{sec:introduction}
Rule-based access control policy models allow flexible configuration of access control policies using a set of crafted rules.
Administrators typically design each rule (or a set of rules) to address a certain protection requirement in the target system.
This is done by defining a condition on the access context that determines the applicability of the rule and the corresponding access decision that should follow.
There are several factors in the design of a rule-based access control model that can affect its expressiveness, i.e., its ability to represent a diverse range of policies.
Design choices such as the support for testing complemented conditions, specifying both \permit and \deny rules, as well as choosing various conflict resolution strategies are generally perceived to improve the expressiveness of a model.
However, their exact impacts on the model expressiveness are unclear.
This usually leads to claims about the expressiveness of the designed models that are hardly verifiable.
Also, a common misconception is that providing additional features and choices in a model is better.
However, we note that providing those features typically comes at the cost of complicating the policies and hence an increased risk of policy misconfigurations by administrators. 
Therefore, it is critical to assess the expressiveness of a model in order to make a more balanced decision about the suitable design choices for a given application setting.
In this work, we systematically investigate the impact of the design choices in a rule-based policy model on its expressiveness.
We propose a general framework that allows formally capturing the semantics of rule-based policies according to the rule structures and compositions as well as characterizing those semantics in terms of well-formedness properties.
Our approach allows an intuitive way of characterizing and comparing the expressiveness of rule-based policy model designs.

Given the multiplicity of the design choices in rule-based policy models that lead to their different expressiveness, we also investigate the policy conversion (or migration) problem in this context. 
The need for policy conversion commonly arises in situations where a system adopts a new policy model and needs to express its currently-enforced policy in this new model.
A system may also need to adopt an external policy instance that has been specified in the context of a different model.
While the expressiveness of two rule-based models could be different and therefore one may not be able to represent all policies expressible in another, particular policy 
\emph{instances} 
may be expressible in both models. Therefore, policy conversion could be possible in such instances without sacrificing the semantics of the policy of interest.
Based on our framework for rule-based policy semantics and well-formedness properties, we formulate and address two problems in this context: whether a policy instance in a source model can be converted to a semantically-equivalent policy instance in a target model, and how to perform the conversion if it is possible.


The work presented in this paper is a comprehensive treatment of the abovementioned problems, extending and generalizing our earlier publications~\cite{MNI21-SACMAT,ruiz2024sacmat}.
The following summarizes our major contributions:
\begin{itemize}
  \item
  We propose a framework of rule-based policy models that capture the common design choices in these models including allowing tests for complemented conditions, permit/deny rules, default decision, and conflict resolution strategies (\Cref{sec:models}).
  We purposefully use an abstract notion of conditions to allow our results to be applicable to a wide range of rule-based policy models including attribute-based access control (ABAC)~\cite{ServosHGABACFormalModel2014,crampton2012ptacl,eXtensibleAccessControlb} and relationship-based access control (ReBAC)~\cite{masoumzadeh2010osnac,fong2011sacmat,pasarella2017datalog}.
  \item
  We propose the notion of \emph{condition minterms} as the fine-grained building blocks of policy semantics and discuss how the semantics of rule condition expressions, and hence policies, can be expressed based on condition minterms (\Cref{sec:semantics}).
  \item
  We define four well-formedness properties for sets of condition minterms: \emph{upward-closedness}, \emph{downward-closedness}, \emph{convexity}, and \emph{co-convexity}.
  We propose how rule expressions, and rule-based policies composed of them, exhibit such well-formedness properties (\Cref{sec:wellformedness}).
  We show this by characterizing the impacts of algebraic operations on well-formed sets of condition minterms, and provide a pictorial approach for an intuitive understanding of those impacts.
  \item
  We reason about the expressive power of several models built using our framework. The expressiveness results characterize the exact set of policy semantics that each model can represent. We also establish an expressiveness hierarchy of the models (\Cref{sec:expressiveness}).
  \item
  We formulate the problems of \emph{policy convertibility} and \emph{policy conversion} in the context of rule-based policy models, and show how our expressiveness framework can be utilized to address these problems (\Cref{sec:conversion}). 
  In particular, we derive the complexity of convertibility problem considering different rule-based models as source and target.
\end{itemize}

In addition to the abovementioned sections, we review the closely related work in \Cref{sec:related} and provide conclusion and future work directions in \Cref{sec:conclusion}.


\section{Framework of Rule-Based Models}
\label{sec:models}
In this section, we present a framework for defining various rule-based access control policy models based on choices on common features such as conditional expressions and rule effects.
At its core, any rule-based policy consists of a set of rules.
Each rule in the policy is specified based on a conditional expression that is a conjunction (product) of a set of conditions or their complements.
We denote the set of all conditions that can be used in a system by \allconditions and the complement of condition $c \in \allconditions$ by $\overline{c}$.
We also consider a special conditional expression \emph{true}, denoted by~$\top$, that is always satisfied.
We denote the set of all possible (conjunctive) conditional expressions by \allexpressions.
We consider an abstract notion of conditions where each condition is evaluated to a truth value based on an access request information at runtime.
Depending on the particular specification and evaluation model of conditions, our results will be applicable to a variety of rule-based policies, such as attribute-based access control (ABAC) and relationship-based access control (ReBAC) policies.
In the context of ABAC, for example, a condition may indicate whether the role of the subject of the access request is student, or whether the sensitivity level of the object of the access request is high.
In the context of ReBAC, for instance, a condition may test whether the access subject owns the folder that contains the access object (i.e., they are related by a sequence of ``owns'' and ``contains'' relationships).
A rule is applicable to an access request, if its conditional expression is satisfied in the context of access request (i.e., based on the truth assignment to conditions in $C$).

In addition to a conditional expression, a rule 
consists of a decision (or effect): either \permit or \deny.
Therefore, a rule can be captured as $\tuple{\phi, d}$, where $\phi \in \allexpressions$ and $d \in \set{\permit, \deny}$.
We use the dot notation to refer to the elements of a rule.
Since both \permit and \deny decisions are possible based on applicable rules to an access request, a \emph{conflict resolution strategy} is used to determine the final access decision.
Two straightforward conflict resolution strategies are \emph{\permit-overrides} and \emph{\deny-overrides}.
Another common strategy is \emph{first-applicable}, where rules are processed in their specified order and the access decision determined by the first applicable rule.
To support such a strategy, we define a policy as an ordered set (tuple) of form $\tuple{r_1, r_2, \ldots, r_n}$.
We denote the size of policy $p$, i.e., the number of rules in the policy, as $|p|$.

Our framework can define a rule-based policy model based on the abovementioned concepts using a 4-tuple \tuple{\alpha, \beta, \delta, \sigma}, where
\begin{itemize}
\item $\alpha \in \set{\textrm{\uncomplementedexp, \complementedexp, \compuncompexp}}$ indicates whether only uncomplemented conditions (\uncomplementedexp), only complemented conditions (\complementedexp), or both (\compuncompexp) are allowed in rule expressions.
\item $\beta \subseteq \set{\permit, \deny}$ indicates whether only \permit rules, only \deny rules, or both are allowed.
\item $\delta \in \set{\permit, \deny}$ indicates the default decision when no rule is applicable to an access request.
\item $\sigma \in \set{\textrm{\deny-overrides}, \textrm{\permit-overrides, \textrm{first-applicable}}}$ indicates the conflict resolution strategy when conflicting \permit and \deny rules apply to an access request.
\end{itemize}

Different rule-based policy models can be constructed depending on the choice of the above parameters.
In the rest of the paper, we focus on six such models defined below.
\begin{description}
\item[Negation] \tuple{\compuncompexp, \set{\permit}, \deny, -}.
\item[DDDO] \tuple{\uncomplementedexp, \set{\permit, \deny}, \deny, \textrm{\deny-overrides}}.
\item[DDPO] \tuple{\uncomplementedexp, \set{\permit, \deny}, \deny, \textrm{\permit-overrides}}.
\item[DPDO] \tuple{\uncomplementedexp, \set{\permit, \deny}, \permit, \textrm{\deny-overrides}}.
\item[DPPO] \tuple{\uncomplementedexp, \set{\permit, \deny}, \permit, \textrm{\permit-overrides}}.
\item[DDFA] \tuple{\uncomplementedexp, \set{\permit, \deny}, \deny, \textrm{first-applicable}}.
\end{description}

The Negation model supports both complemented and uncomplemented conditions, but only \permit rules.
Note that the choice of $\sigma$ does not matter in this model  (shown as ``$-$'') as there will be no conflicting decisions.
The rest of the models above such as DDDO (short for \emph{default-deny and deny-overrides}) and DPDO (short for \emph{default-permit and deny-overrides}) support only uncomplemented conditions but both \permit and \deny rules.
We also note that the list above is not exhaustive (e.g., misses \compuncompexp models that allow both types of rule decisions).
However, the list is representative of the models explored in prior literature and adequately demonstrates the strengths of the contributions in this paper.

The following examples show two small policies based on the Negation and DDDO models.
We later show (in~\Cref{sec:semantics}) that these two policies are semantically equivalent).
\begin{example}[Negation Policy]
  \label{ex:negation-policy}
  The following Negation policy permits accesses
  when $c_1$ is satisfied and $c_3$ is not satisfied,
  or when $c_2$ is satisfied and $c_3$ is not satisfied.
  Access is denied in all other cases.
  \begin{policy}
  \item $\tuple{c_1\overline{c_3}, \permit}$
  \item $\tuple{c_2\overline{c_3}, \permit}$
  \end{policy}

\end{example}
\begin{example}[DDDO Policy]
  \label{ex:dddo-policy}
  The following DDDO policy permits access when $c_1$ or $c_2$ is satisfied.
  However, it will deny access when $c_3$ is satisfied
  (\emph{irrespective of whether $c_1$ or $c_2$ is
  satisfied}).
  Note the potential conflict between the two cases, and that \deny overrides in this model.
  \begin{policy}
  \item $\tuple{c_1, \permit}$
  \item $\tuple{c_2, \permit}$
  \item $\tuple{c_3, \deny}$
  \end{policy}
\end{example}

\section{Policy Semantics}
\label{sec:semantics}
In this section, we present a universal approach to capture the semantics of rule expressions and rule-based policies defined based on the framework discussed in \Cref{sec:models}.
\subsection{Condition Minterms}
\label{sec:minterm}
Every condition is essentially a boolean variable, evaluating to either true or false.
Considering the set of all possible conditions (\allconditions), we can define the notion of \emph{condition minterm}:
\begin{definition}[Condition Minterm]
  A \emph{condition minterm}, or a \emph{minterm} for short, is a conditional expression where exactly one instance of each condition or its complement is present.
  Formally, a condition minterm is constructed using $\prod_{c_i \in \allconditions}(c_i | \overline{c_i})$ where ``$\mid$'' is a choice operator.
\end{definition}
We denote the set of all minterms by \allminterms.
Since all conditions are present in a minterm, in either their direct (uncomplemented) or complemented form,
there are $2^{|\allconditions|}$ possible minterms.
Note that, for each minterm, only one combination of truth values for conditions makes that minterm true.
Therefore, we can use minterms as the unit of capturing policy semantics.

We define a partial order relation ($\le$) on condition minterms based on their uncomplemented conditions as follows:
\begin{definition}[Partial Ordering of Minterms]
  \label{def:minterm-poset}
  Let $S(m)$ be the set of uncomplemented conditions in condition minterm $m$.
  For minterms $m_1$ and $m_2$, $m_1 \le m_2$ iff $S(m_1) \subseteq S(m_2)$.
\end{definition}
We can extend the above definition to a \emph{strict ordering}: $m_1 < m_2$ iff $m_1 \le m_2$ and $m_1 \neq m_2$. 

\subsection{Expression and Policy Semantics}
\label{sec:policy-minterms}
When an access control request is evaluated, each condition in the system is evaluated as either true or false.
That renders one of the condition minterms in \allminterms as true, and the rest of them as false.
Based on this property, each condition minterm can be considered as one possible situation in a system.
Thus, we use condition minterms to express the semantics of conditional expressions and policies.
The semantics of a conditional expression determines in what situations the expression is evaluated as true:
\begin{definition}[Expression-Minterms]
  \label{def:expression-minterms}
  Given conditional expression $\phi$, its \emph{expression-minterms}, denoted by $\mu(\phi)$, is defined as the set of minterms whose disjunction is logically equivalent to $\phi$.
  Let $\allconditions_\phi$ be the set of conditions present in $\phi$.
  Formally, $\mu(\phi) = \phi \circ \allminterms[\allconditions \smallsetminus \allconditions_\phi]$.
\end{definition}

In the above definition, $\allminterms[\allconditions \smallsetminus \allconditions_\phi]$ represents the set of minterms produced by conditions that are not in $\phi$. 
Operator ``$\circ$'' produces minterms (that belong to \allminterms) as the conjunction of~$\phi$ and other conditions not in $\phi$ (in their uncomplemented or complemented form).

\begin{example}[Expression-Minterms]
\label{ex:negation-policy-rule-minterms}
  Given $\allconditions = \set{c_1, c_2, c_3}$, the expression-minterms of the rules in the policy in \Cref{ex:negation-policy} are 
  $\set{c_1\overline{c_2}\overline{c_3}, c_1c_2\overline{c_3}}$ 
  and $\set{\overline{c_1}c_2\overline{c_3}, c_1c_2\overline{c_3}}$, 
  respectively.
\end{example}

\begin{example}[Expression-Minterms]
\label{ex:dddo-policy-rule-minterms}
  In the case of \Cref{ex:dddo-policy}, given $\allconditions = \set{c_1, c_2, c_3}$, the expression-minterms of its rules are
  $\set{c_1\overline{c_2}\overline{c_3}, c_1\overline{c_2}c_3, c_1c_2\overline{c_3}, c_1c_2c_3}$,
  $\set{\overline{c_1}c_2\overline{c_3}, \overline{c_1}c_2c_3, c_1c_2\overline{c_3}, c_1c_2c_3}$,
  and $\set{\overline{c_1}\overline{c_2}c_3, \overline{c_1}c_2c_3, c_1\overline{c_2}c_3, c_1c_2c_3}$, respectively.  
\end{example}

Rule-based policies permit or deny access requests based on the evaluation results of the conditional expressions in their rules.
The semantics of such policies can be expressed based on the situations (i.e., condition minterms) in which they permit access requests, defined as follows (slightly abusing the $\mu$ notation used earlier):
\begin{definition}[Policy-Minterms]
  \label{def:policy-minterms}
  Given policy $p$, its \emph{policy-minterms}, denoted by $\mu(p)$, is defined as the set of minterms where $p$ permits an access request iff one of those minterms is true.
\end{definition}

\begin{table}
  \centering
  \caption{Policy Model Semantics Calculated Based on Rule Semantics.}
  \label{tab:model-semantics}
  \addtolength\tabcolsep{-1pt}
  \renewcommand{\arraystretch}{1.5}
  \begin{tabular}{c|p{2.7in}}
    \hline
    \boldmath\textbf{Model} & \boldmath\textbf{Semantics of Policy $p$} \\
    \hline
    \hline
    Negation & $\bigcup_{r \in p}{\mu(r.\phi)}$ \\
    \hline
    DDDO & $\bigcup_{r \in p \, \land \, r.d = \permit}{\mu(r.\phi)} \; \smallsetminus \; \bigcup_{r \in p \, \land \, r.d = \deny}{\mu(r.\phi)}$ \\
    \hline
    DPPO & $\allminterms \smallsetminus \bigcup_{r \in p \, \land \, r.d = \deny}{\mu(r.\phi)} \cup \bigcup_{r \in p \, \land \, r.d = \permit}{\mu(r.\phi)}$ \\
    \hline
    DDPO & $\bigcup_{r \in p \, \land \, r.d = \permit}{\mu(r.\phi)}$ \\
    \hline
    DPDO & $\allminterms \smallsetminus \bigcup_{r \in p \, \land \, r.d = \deny}{\mu(r.\phi)}$ \\
    \hline
    DDFA &
           \vspace{-1em}
           \begin{algorithmic}
             \State $M \gets \emptyset$
             \For{$r$ in $p$ iterated in reverse order}
               \If{$r.d = \permit{}$}
                 \State $M \gets M \cup \mu(r.\phi)$
               \ElsIf{$r.d = \deny{}$}
                 \State $M \gets M \smallsetminus \mu(r.\phi)$
               \EndIf
             \EndFor
             \State $\mu(p) \gets M$
           \end{algorithmic}
    \\
    \hline
  \end{tabular}
\end{table}

The policy-minterms for a policy can be derived from the expression-minterms and decisions of its rules based on its policy model.
\Cref{tab:model-semantics} presents how such semantics can be calculated depending on the policy model used.
As shown, the calculations involve simple set operations (union and subtraction) on expression-minterms of policy rules.
For instance, the semantics of a DDDO policy can be obtained by subtracting the union of the expression-minterms of its \deny rules from the union of the expression-minterms of its \permit rules.
Recall that \allminterms is the set of all condition minterms, which acts as the starting semantics for policy models with \permit as their default decision.
The semantics of the DDFA model is slightly more involved compared to other models since expression-minterms of rules need to be considered in order.

\begin{example}[Policy-Minterms and Semantic Equivalence]
Using the formulae in \Cref{tab:model-semantics}, the Negation policy in \Cref{ex:negation-policy} and the DDDO policy in \Cref{ex:dddo-policy} can be shown to be semantically equivalent.
In the case of \Cref{ex:negation-policy}, its policy-minterms formed by the union of its rules' expression-minterms shown in \Cref{ex:negation-policy-rule-minterms} will be $\set{\overline{c_1}c_2\overline{c_3}, c_1\overline{c_2}\overline{c_3}, c_1c_2\overline{c_3}}$.
In the case of \Cref{ex:dddo-policy}, given the expression-minterms of its rules shown in \Cref{ex:dddo-policy-rule-minterms}, we need to subtract the expression-minterms of the third (\deny) rule from the union of expression-minterms of the first two (\permit) rules, which will result in the same final policy-minterms.
\end{example}

\section{Well-Formed Policy Semantics}
\label{sec:wellformedness}
In this section, we show that the semantics of policies in different models follow certain well-formedness properties, which are helpful in the analysis of policy models.
We begin with defining these well-formedness properties for sets of condition minterms.
Then, we show how the semantics of rule expressions are well-formed, and how we can reason about the well-formedness of the semantics of different policy models according to how they interpret their policy rules.

\subsection{Well-Formed Sets of Minterms}
\label{sec:wellformed-sets}
Based on the partial order relation on minterms, we define four well-formedness properties for sets of minterms:
\begin{definition}[Upward-Closed Set of Minterms]
  The upward closure of set of minterms $M$, denoted by $\upwardclosure{M}$,
  is $\set{u \,|\, \exists m \in M: m \le u}$.
  $M$ is \emph{upward-closed} iff $\upwardclosure{M} = M$.
\end{definition}

\begin{definition}[Downward-Closed Set of Minterms]
  The downward closure of set of minterms $M$, denoted by $\downwardclosure{M}$,
  is $\set{u \,|\, \exists m \in M: u \le m}$.
  $M$ is \emph{downward-closed} iff $\downwardclosure{M} = M$.
\end{definition}

\begin{definition}[(Co-)Convex Set of Minterms]
  Set of minterms $M$ is \emph{convex} iff for every pair of minterms $m_1 \le m_2$ in $M$: $\set{m \,|\, m_1 \le m \le m_2} \subseteq M$.
  Set of minterms $M$ is \emph{co-convex} iff its complement ($\allminterms \smallsetminus M$) is convex.
\end{definition}

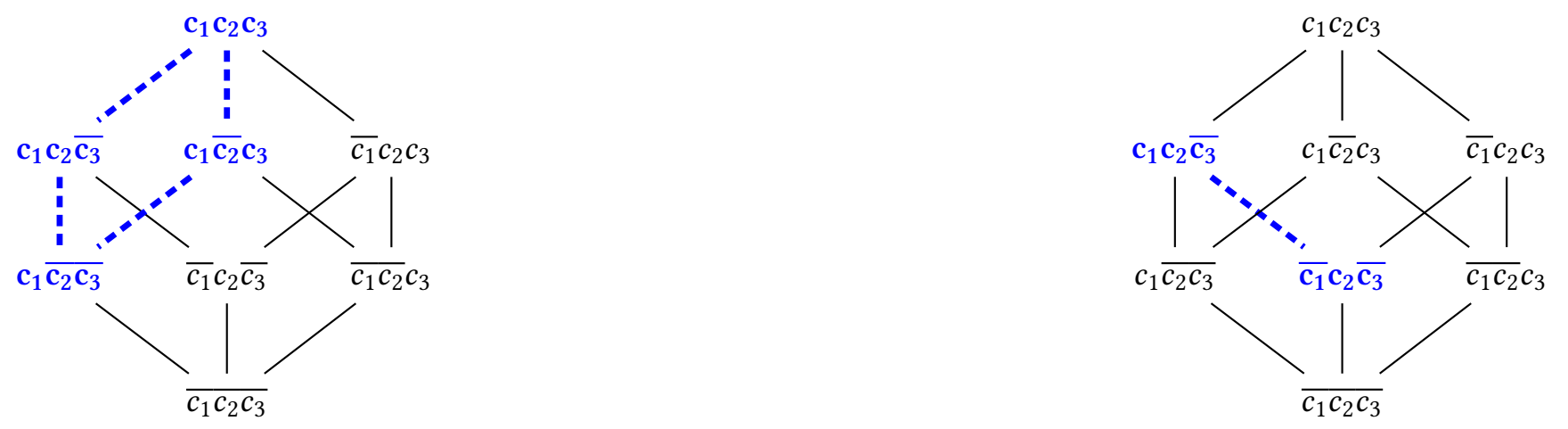
\begin{figure}
  \centering
  \begin{subfigure}[t]{0.45\columnwidth}
    \centering
  \adjustbox{scale=0.8,center}{%
  \begin{tikzcd}[tips=false,column sep=1.5em,row sep=2.1em]
    & \hasseFont{c_1c_2c_3} \arrow[dl, hasse] \arrow[d, hasse] \arrow[dr] & \\
    \hasseFont{c_1c_2\overline{c_3}} \arrow[d, hasse] \arrow[dr] & \hasseFont{c_1\overline{c_2}c_3} \arrow[dl, hasse] \arrow[dr] & \overline{c_1}c_2c_3 \arrow[dl] \arrow[d]\\
    \hasseFont{c_1\overline{c_2}\overline{c_3}} \arrow[dr] & \overline{c_1}c_2\overline{c_3} \arrow[d] & \overline{c_1}\overline{c_2}c_3 \arrow[dl] \\
    & \overline{c_1}\overline{c_2}\overline{c_3} &
  \end{tikzcd}}
  \caption{Upward-closed set of minterms: \set{c_1c_2c_3, c_1c_2\overline{c_3}, c_1\overline{c_2}c_3, c_1\overline{c_2}\overline{c_3}}.}
    \label{fig:example.upward_closed}
  \end{subfigure}
\hfill
  \begin{subfigure}[t]{0.45\columnwidth}
    \centering
    \adjustbox{scale=0.8,center}{%
      \begin{tikzcd}[tips=false,column sep=1.5em,row sep=2.1em]
      & c_1c_2c_3 \arrow[dl] \arrow[d] \arrow[dr] & \\
      \hasseFont{c_1c_2\overline{c_3}} \arrow[d] \arrow[dr, hasse] & c_1\overline{c_2}c_3 \arrow[dl] \arrow[dr] & \overline{c_1}c_2c_3 \arrow[dl] \arrow[d]\\
      c_1\overline{c_2}\overline{c_3} \arrow[dr] & \hasseFont{\overline{c_1}c_2\overline{c_3}} \arrow[d] & \overline{c_1}\overline{c_2}c_3 \arrow[dl] \\
      & \overline{c_1}\overline{c_2}\overline{c_3} &
      \end{tikzcd}
    }
  \caption{Convex set of minterms: \set{c_1c_2\overline{c_3}, \overline{c_1}c_2\overline{c_3}}.}
  \label{fig:example.convex}
  \end{subfigure}
  \caption{Well-formed sets of condition minterms. Bold, blue text indicate set elements.}
  \label{fig:hasse-examples}
\end{figure}

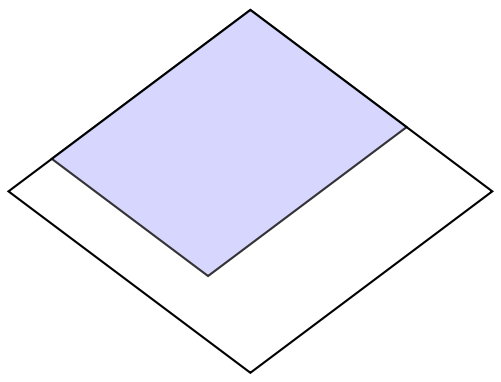
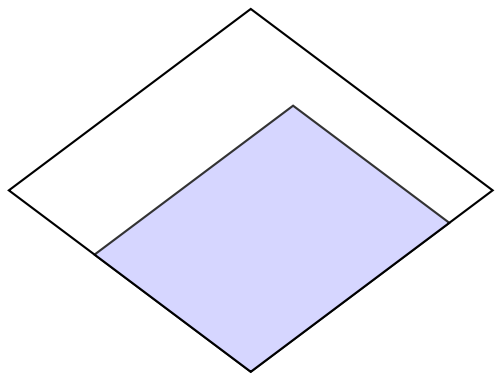
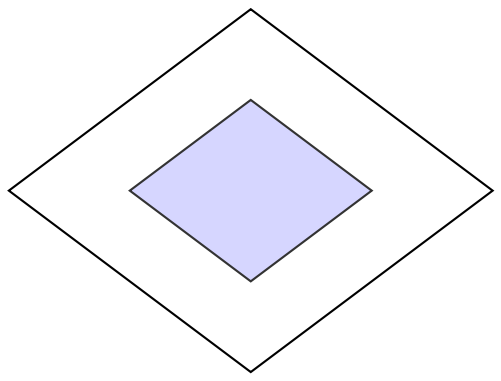
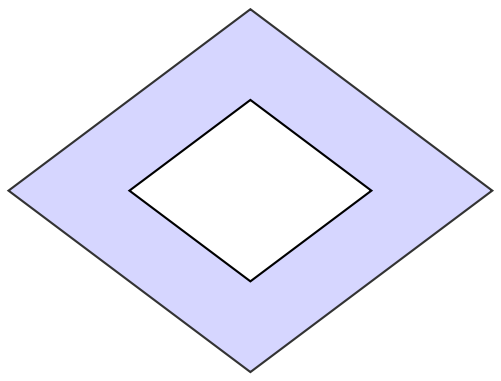
\begin{figure}
  \centering
  \begin{subfigure}[t]{0.45\columnwidth}
    \centering
\begin{tikzpicture}[scale=0.4]
  \draw (4,0) -- (0,3) -- (-4,0) -- (0,-3) -- cycle;
  \path[name path=r1] (4,0) -- (0,3);
  \path[name path=r2] (0,3) -- (-4,0);
  \path[name path=t1] (-0.7, -1.4) -- (-4.7,1.6);
  \path[name path=t2] (-0.7, -1.4) -- (3.3,1.6);
  \draw [name intersections = {of=t1 and r2, by = x}] (x);
  \draw [name intersections = {of=t2 and r1, by = y}] (y);
  \fill[permit] (-0.7, -1.4) -- (x) -- (0,3) -- (y) -- cycle node[xshift=-0.3cm, yshift=0.3cm, thick, black] {};
\end{tikzpicture}
\caption{Upward-closed set of minterms.}
    \label{fig:minterm.upward-closed}
  \end{subfigure}
\hfill
  \begin{subfigure}[t]{0.45\columnwidth}
    \centering
\begin{tikzpicture}[scale=0.4]
  \draw (-4,0) -- (0,-3) -- (4,0) -- (0,3) -- cycle;
  \path[name path=r1] (-4,0) -- (0,-3);
  \path[name path=r2] (0,-3) -- (4,0);
  \path[name path=t1] (0.7, 1.4) -- (4.7,-1.6);
  \path[name path=t2] (0.7, 1.4) -- (-3.3,-1.6);
  \draw [name intersections = {of=t1 and r2, by = x}] (x);
  \draw [name intersections = {of=t2 and r1, by = y}] (y);
  \fill[permit] (0.7, 1.4) -- (x) -- (0,-3) -- (y) -- cycle node[xshift=-0.3cm, yshift=0.3cm, thick, black] {};
\end{tikzpicture}
\caption{Downward-closed set of minterms.}
    \label{fig:minterm.downward-closed}
  \end{subfigure}  
\begin{subfigure}[t]{0.45\columnwidth}
    \centering
\begin{tikzpicture}[scale=0.4]
  \draw (4,0) -- (0,3) -- (-4,0) -- (0,-3) -- cycle;
  \fill[permit] (2,0) -- (0,1.5) -- (-2,0) -- (0,-1.5) -- cycle node[xshift=-0.3cm, yshift=0cm, thick, black] {};
\end{tikzpicture}
\caption{Convex set of minterms.}
    \label{fig:minterm.convex}
  \end{subfigure}
\hfill
  \begin{subfigure}[t]{0.45\columnwidth}
    \centering
\begin{tikzpicture}[scale=0.4]
  \draw[permit] (4,0) -- (0,3) -- (-4,0) -- (0,-3) -- cycle;
  \fill[white, draw=black] (2,0) -- (0,1.5) -- (-2,0) -- (0,-1.5) -- cycle node[xshift=-0.3cm, yshift=0cm, thick, black] {};
\end{tikzpicture}
\caption{Co-convex set of minterms.}
    \label{fig:minterm.coconvex}
  \end{subfigure}  
  \caption{Visualizing well-formed sets of minterms.
  Outer diamond-shaped space resembles a Hasse diagram corresponding to all condition minterms as shown in \Cref{fig:hasse-examples}. 
  Filled (blue) areas represent various well-formed minterms sets.}
  \label{fig:minterm}
\end{figure}

We illustrate the above well-formedness properties with the help of the Hasse diagram corresponding to the condition minterms for $\allconditions{} = \set{c_1, c_2, c_3}$ in \Cref{fig:hasse-examples}.
\Cref{fig:example.upward_closed} highlights an upward-closed subset of minterms (using bold, blue text and connecting dashed lines).
This set of minterms is also convex.
Its complement (non-highlighted minterms) is also convex and downward-closed.
\Cref{fig:example.convex} highlights a convex subset of minterms.
Here, the complement of the highlighted set of minterms is not convex.
To facilitate visual inspection of well-formedness properties more easily, we propose a set of abstract shapes in \Cref{fig:minterm} that resemble highlighted minterm sets in a Hasse diagram.
In particular, \Cref{fig:minterm.upward-closed,fig:minterm.downward-closed,fig:minterm.convex,fig:minterm.coconvex} illustrate sets of minterms that are upward-closed, downward-closed, convex, and co-convex, respectively.

We note that the well-formedness properties are not mutually exclusive.
In particular, we capture the specificity/generalization relationships between well-formed sets of minterms in the following lemma:
\begin{lemma}
  \label{lem:wellformed-sets-partial-order}
  The following statements are correct about well-formed sets of minterms:
  \begin{thmenum}
  \item 
    Sets $\emptyset$ and {\allminterms} are both upward-closed and downward-closed.
  \item
    Upward-closed sets are (co-)convex.
  \item
    Downward-closed sets are (co-)convex.
  \end{thmenum}  
\end{lemma}

Note that convexity and co-convexity are the most general well-formedness properties that we consider in this paper.

\subsection{Well-formedness of Rule Expressions}
\label{sec:wellformed-expressions}
Interestingly, the semantics of rule expressions in our framework (\Cref{sec:models}) are well-formed and can be further characterized depending on their use of (un)complemented conditions.
\begin{lemma}
  \label{lem:wellformed-rules}
  The semantics of \uncomplementedexp, \complementedexp, and \compuncompexp rule expressions are upward-closed, downward-closed, and convex, respectively.
\end{lemma}

\begin{proof}[Proof of \uncomplementedexp semantics well-formedness]
  Let $\phi$ be an \uncomplementedexp expression, $\allconditions_\phi$ be the set of conditions present in $\phi$, and $m_1$ be any minterm in $\mu(\phi)$.
  By construction (\Cref{def:expression-minterms}), $m_1$ has to be of form $\phi m_1'$, where $m_1' \in \allminterms[\allconditions \smallsetminus \allconditions_\phi]$.
  Let $S()$ denote the set of uncomplemented conditions in a minterm.
  Clearly, $\allconditions_\phi \subseteq S(m_1)$.
  For any minterm $m_2 \ge m_1$, we should have $S(m_2)\supseteq S(m_1) \supseteq \allconditions_\phi$.
  Therefore, $m_2$ has to be of form $\phi m_2'$.
  Furthermore, $m_2'$ has to belong to $\allminterms[\allconditions \smallsetminus \allconditions_\phi]$.
  Thus $m_2 \in \mu(\phi)$, proving that $\mu(\phi)$ is upward-closed.
\end{proof}

\begin{proof}[Proof of \complementedexp semantics well-formedness]
  Let $\phi$ be an \complementedexp expression, $\allconditions_\phi$ be the set of conditions present in $\phi$, and $m_1$ be any minterm in $\mu(\phi)$.
  By construction (\Cref{def:expression-minterms}), $m_1$ has to be of form $\phi m_1'$, where $m_1' \in \allminterms[\allconditions \smallsetminus \allconditions_\phi]$.
  As before, let $S()$ denote the set of uncomplemented conditions in a minterm.
  Since all conditions in $\phi$ are complemented, $S(m_1) = S(m_1')$.
  For any minterm $m_2 \le m_1$, we should have $S(m_2) \subseteq S(m_1) = S(m_1')$.
  Therefore, $m_2$ has to be of form $\phi m_2'$.
  Furthermore, $m_2'$ has to belong to $\allminterms[\allconditions \smallsetminus \allconditions_\phi]$.
  Thus $m_2 \in \mu(\phi)$, proving that $\mu(\phi)$ is downward-closed.
\end{proof}

\begin{proof}[Proof of \compuncompexp semantics well-formedness]
  Let $\phi$ be a \compuncompexp expression, $\allconditions_\phi$ be the set of conditions present in $\phi$, and $m_1, \, m_2, \, m_3$ be minterms such that~$m_1 \, < \, m_2 \, < \, m_3$
  where $m_1, \, m_3 ~ \in ~ \mu( \phi )$.
  Clearly, $\phi$ is a \emph{subexpression} of both $m_1$ and $m_3$~(by Definition~\ref{def:expression-minterms}), i.e.,
  $m_1$ has to be of form $\phi m_1'$, where $m_1' \in \allminterms[\allconditions \smallsetminus \allconditions_\phi]$,
  and similarly, 
  $m_3$ has to be of form $\phi m_3'$, where $m_3' \in \allminterms[\allconditions \smallsetminus \allconditions_\phi]$.
Thus  they can differ
  only on variables (conditions) that \emph{do not} occur in~$\phi$.
  Then $\phi$ is also a subexpression of~$m_2$ and
  we are done by~Definition~\ref{def:expression-minterms}.
\end{proof}

\subsection{Policy Shapes}
\label{sec:wellformed-policies}


\begin{table}
  \centering
  \addtolength\tabcolsep{-2pt}
  \caption{Operations on Well-Formed Sets of Minterms ($\emptyset$: Empty Set of Minterms; \allminterms: Set of All Minterms; UC: Upward-Closed; DC: Downward-Closed; CX: Convex; CC: Co-Convex; ``-'': Not Necessarily Well-Formed).}
  \label{tab:wellformed-operations}
  \subfloat[Union]{
    \begin{tabular}{c|cccccc}
      $\cup$       & $\emptyset$  & \allminterms & UC           & DC           & CX           & CC           \\
      \hline                                                                                                
      $\emptyset$  & $\emptyset$  & \allminterms & UC           & DC           & CX           & CC           \\
      \allminterms & \allminterms & \allminterms & \allminterms & \allminterms & \allminterms & \allminterms \\
      UC           & UC           & \allminterms & UC           & CC           & -            & CC           \\
      DC           & DC           & \allminterms & CC           & DC           & -            & CC           \\
      CX           & CX           & \allminterms & -            & -            & -            & -           \\
      CC           & CC           & \allminterms & CC           & CC           & -           & CC           \\
    \end{tabular}
    \label{tab:wellformed-operations.union}
  }
  \hspace{1.5cm}
  \subfloat[Subtraction]{
    \begin{tabular}{c|cccccc}
      $\smallsetminus$  & $\emptyset$  & \allminterms & UC          & DC          & CX          & CC          \\
      \hline                                                         
      $\emptyset$  & $\emptyset$  & $\emptyset$  & $\emptyset$ & $\emptyset$ & $\emptyset$ & $\emptyset$ \\
      \allminterms & \allminterms & $\emptyset$  & DC          & UC          & CC          & CX          \\
      UC           & UC           & $\emptyset$  & CX          & UC          & -           & CX          \\
      DC           & DC           & $\emptyset$  & DC          & CX          & -           & CX          \\
      CX           & CX           & $\emptyset$  & CX          & CX          & -           & CX          \\
      CC           & CC           & $\emptyset$  & -          & -          & -           & -           \\
    \end{tabular}
    \label{tab:wellformed-operations.subtraction}
  }
\end{table}


As discussed in \Cref{sec:policy-minterms}, the semantics of policies can be derived using simple set operations on the semantics of their rules.
We show in \Cref{tab:wellformed-operations} that such set operations, specifically, union (\Cref{tab:wellformed-operations.union}) and subtraction (\Cref{tab:wellformed-operations.subtraction}), could also induce well-formedness properties when operating on well-formed sets of minterms.
For instance, subtracting a downward-closed set from an upward-closed set (UC $\smallsetminus$ DC) results in an upward-closed set.
As another example, the result of subtraction of a convex set from an upward-closed set is not necessarily well-formed. 
As we showed in~\Cref{lem:wellformed-sets-partial-order}, there is a specificity relationship between certain well-formedness properties.
For example, an upward-closed set is also convex (but not vice versa).
The results in \Cref{tab:wellformed-operations} are most helpful when we consider the most specific well-formedness designation of the operands.
The row\slash column corresponding to upward-closed and downward-closed sets (UC and DC, respectively) should be considered only if the operand is neither $\emptyset$ nor \allminterms.
Similarly, the row\slash column corresponding to convex sets (CX) should be considered only if the operand is neither upward-closed nor downward-closed.


\begin{figure}
  \centering
  \begin{subfigure}[t]{0.49\columnwidth}
    \centering
\begin{tikzpicture}[scale=0.5]
  \draw (4,0) -- (0,3) -- (-4,0) -- (0,-3) -- cycle;
  \path[name path=r1] (4,0) -- (0,3);
  \path[name path=r2] (0,3) -- (-4,0);
  \path[name path=t1] (-0.7, -1.4) -- (-4.7,1.6);
  \path[name path=t2] (-0.7, -1.4) -- (3.3,1.6);
  \draw [name intersections = {of=t1 and r2, by = x}] (x);
  \draw [name intersections = {of=t2 and r1, by = y}] (y);
  \fill[permit] (-0.7, -1.4) -- (x) -- (0,3) -- (y) -- cycle node[xshift=-0.3cm, yshift=0.3cm, thick, black] {$\mathit{UC_1}$};
  \path[name path=t3] (1.5, -1) -- (-2.5,2);
  \path[name path=t4] (1.5, -1) -- (5.5,2);
  \draw [name intersections = {of=t3 and r2, by = w}] (w);
  \draw [name intersections = {of=t4 and r1, by = z}] (z);
  \fill[permit] (1.5, -1) -- (w) -- (0,3) -- (z) -- cycle node[xshift=-0.1cm, yshift=0.9cm, thick, black] {$\mathit{UC_2}$};
\end{tikzpicture}

\caption{$\mathit{UC_1 \cup UC_2}$ is upward-closed.}
  \end{subfigure}
\hfill
\begin{subfigure}[t]{0.49\columnwidth}
\centering
\begin{tikzpicture}[scale=0.5]
  \draw (4,0) -- (0,3) -- (-4,0) -- (0,-3) -- cycle;
  \path[name path=r1] (4,0) -- (0,3);
  \path[name path=r2] (0,3) -- (-4,0);
  \path[name path=t1] (-0.7, -1.4) -- (-4.7,1.6);
  \path[name path=t2] (-0.7, -1.4) -- (3.3,1.6);
  \draw [name intersections = {of=t1 and r2, by = x}] (x);
  \draw [name intersections = {of=t2 and r1, by = y}] (y);
  \fill[permit] (-0.7, -1.4) -- (x) -- (0,3) -- (y) -- cycle node[xshift=-0.3cm, yshift=0.3cm, thick, black] {$\mathit{UC_1}$};
  \path[name path=t3] (1.5, -1) -- (-2.5,2);
  \path[name path=t4] (1.5, -1) -- (5.5,2);
  \draw [name intersections = {of=t3 and r2, by = w}] (w);
  \draw [name intersections = {of=t4 and r1, by = z}] (z);
  \fill[deny] (1.5, -1) -- (w) -- (0,3) -- (z) -- cycle node[xshift=-0.1cm, yshift=0.9cm, thick, black] {$\mathit{UC_2}$};
\end{tikzpicture}
\caption{$\mathit{UC_1 \smallsetminus UC_2}$ is convex, but not upward-closed.}
  \end{subfigure}
\par\bigskip
\begin{subfigure}[t]{0.49\columnwidth}
\centering
\begin{tikzpicture}[scale=0.5]
  \draw (4,0) -- (0,3) -- (-4,0) -- (0,-3) -- cycle;
  \fill[permit] (1,-0.9) -- (2.6,0.3) -- (1,1.5) -- (-0.6,0.3) -- cycle node[xshift=0.3cm, yshift=0.4cm, thick, black] {$\mathit{CX_2}$};
  \fill[permit] (-1,-1.7) -- (0.2,-0.8) -- (-1,0.1) -- (-2.2,-0.8) -- cycle node[xshift=-0.25cm, thick, black] {$\mathit{CX_1}$};
  \draw[dashed, color=red] (0,-3) to[out=100,in=-100] (0,3);
\end{tikzpicture}
\caption{$\mathit{CX_1 \cup CX_2}$ is not necessarily well-formed.}
\label{fig:minterm-union-cx-cx}
  \end{subfigure}
\hfill
\begin{subfigure}[t]{0.49\columnwidth}
\centering
\begin{tikzpicture}[scale=0.5]
  \draw (4,0) -- (0,3) -- (-4,0) -- (0,-3) -- cycle;
  \fill[permit] (-0.5,-2.1) -- (1.9,-0.3) -- (-0.5,1.5) -- (-2.9,-0.3) -- cycle node[xshift=-0.3cm, yshift=0.3cm, thick, black] {$\mathit{CX_1}$};
  \fill[deny] (0.7,-0.4) -- (2.3,0.8) -- (0.7,2.0) -- (-0.9,0.8) -- cycle node[xshift=0.2cm, yshift=0.5cm, thick, black] {$\mathit{CX_2}$};
  \draw[dashed, color=red] (0,-3) to[out=110,in=-110] (0,3);
\end{tikzpicture}
\caption{$\mathit{CX_1 \smallsetminus CX_2}$ is not necessarily well-formed.}
  \end{subfigure}
\par\bigskip
\begin{subfigure}[t]{0.49\columnwidth}
\centering
\begin{tikzpicture}[scale=0.5]
  \draw (4,0) -- (0,3) -- (-4,0) -- (0,-3) -- cycle;
  \path[name path=r1] (4,0) -- (0,3);
  \path[name path=r2] (0,3) -- (-4,0);
  \path[name path=t1] (-0.7, -1.4) -- (-4.7,1.6);
  \path[name path=t2] (-0.7, -1.4) -- (3.3,1.6);
  \draw [name intersections = {of=t1 and r2, by = x}] (x);
  \draw [name intersections = {of=t2 and r1, by = y}] (y);
  \fill[permit] (-0.7, -1.4) -- (x) -- (0,3) -- (y) -- cycle node[xshift=-0.3cm, yshift=0.8cm, thick, black] {$\mathit{UC}$};
  \fill[deny] (1.2,-1.2) -- (2.8,0) -- (1.2,1.2) -- (-0.4,0) -- cycle node[xshift=0.3cm, yshift=0.3cm, thick, black] {$\mathit{CX}$};
  \draw[dashed, color=red] (0,-3) to[out=85,in=-85] (0,3);
\end{tikzpicture}
\caption{$\mathit{UC \smallsetminus CX}$ is not necessarily well-formed.}
  \end{subfigure}
\hfill
\begin{subfigure}[t]{0.49\columnwidth}
\centering
\begin{tikzpicture}[scale=0.5]
  \draw (4,0) -- (0,3) -- (-4,0) -- (0,-3) -- cycle;
  \fill[permit] (-0.5,-2.1) -- (1.9,-0.3) -- (-0.5,1.5) -- (-2.9,-0.3) -- cycle node[xshift=-0.3cm, yshift=0.3cm, thick, black] {$\mathit{CX}$};
  \path[name path=r1] (4,0) -- (0,3);
  \path[name path=r2] (0,3) -- (-4,0);
  \path[name path=t3] (1.5, -1) -- (-2.5,2);
  \path[name path=t4] (1.5, -1) -- (5.5,2);
  \draw [name intersections = {of=t3 and r2, by = w}] (w);
  \draw [name intersections = {of=t4 and r1, by = z}] (z);
  \fill[deny] (1.5, -1) -- (w) -- (0,3) -- (z) -- cycle node[xshift=-0.1cm, yshift=0.9cm, thick, black] {$\mathit{UC}$};
\end{tikzpicture}
\caption{$\mathit{CX \smallsetminus UC}$ is convex.}
  \end{subfigure}
\par\bigskip
\begin{subfigure}[t]{0.49\columnwidth}
\centering
\begin{tikzpicture}[scale=0.5]
  \draw (4,0) -- (0,3) -- (-4,0) -- (0,-3) -- cycle;
  \path[name path=r1] (4,0) -- (0,3);
  \path[name path=r2] (0,3) -- (-4,0);
  \path[name path=t1] (-1.3, -0.1) -- (-4.9,2.6);
  \path[name path=t2] (-1.3, -0.1) -- (2.3,2.6);
  \draw [name intersections = {of=t1 and r2, by = x}] (x);
  \draw [name intersections = {of=t2 and r1, by = y}] (y);
  \fill[permit] (-1.3, -0.1) -- (x) -- (0,3) -- (y) -- cycle node[xshift=-0.1cm, yshift=0.7cm, thick, black] {$\mathit{UC}$};
  \path[name path=r3] (-4,0) -- (0,-3);
  \path[name path=r4] (0,-3) -- (4,0);
  \path[name path=t3] (1.3, 0.1) -- (4.9,-2.6);
  \path[name path=t4] (1.3, 0.1) -- (-2.3,-2.6);
  \draw [name intersections = {of=t3 and r4, by = x}] (x);
  \draw [name intersections = {of=t4 and r3, by = y}] (y);
  \fill[permit] (1.3, 0.1) -- (x) -- (0,-3) -- (y) -- cycle node[xshift=0.1cm, yshift=-0.7cm, thick, black] {$\mathit{DC}$};
  \draw[dashed, color=red] (0,-3) to[out=90,in=-90] (0,3);
\end{tikzpicture}
\caption{$\mathit{UC \cup DC}$ is co-convex.}
\label{fig:minterm-union-uc-dc}
  \end{subfigure}
\hfill
\begin{subfigure}[t]{0.49\columnwidth}
\centering
\begin{tikzpicture}[scale=0.5]
  \draw [permit] (4,0) -- (0,3) -- (-4,0) -- (0,-3) -- cycle;
  \fill[white, draw=black] (2,0) -- (0,1.5) -- (-2,0) -- (0,-1.5) -- cycle node[xshift=-1.2cm, yshift=0.6cm, thick, black] {$\mathit{CC}$};
  \path[name path=r3] (-4,0) -- (0,-3);
  \path[name path=r4] (0,-3) -- (4,0);
  \path[name path=t3] (1.3, 0.1) -- (4.9,-2.6);
  \path[name path=t4] (1.3, 0.1) -- (-2.3,-2.6);
  \draw [name intersections = {of=t3 and r4, by = x}] (x);
  \draw [name intersections = {of=t4 and r3, by = y}] (y);
  \fill[deny] (1.3, 0.1) -- (x) -- (0,-3) -- (y) -- cycle node[xshift=0.1cm, yshift=-0.7cm, thick, black] {$\mathit{DC}$};
  \draw[dashed, color=red] (0,-3) to[out=120,in=-120] (0,3);
\end{tikzpicture}
\caption{$\mathit{CC \smallsetminus DC}$ is not necessarily well-formed.}
  \end{subfigure}
  
  \caption{Visual demonstration of operations on well-formed sets of minterms. 
  $\mathit{UC_i}$ is upward-closed, $\mathit{DC}$ is downward-closed, $\mathit{CX_j}$ is convex, and $\mathit{CC}$ is co-convex.
  The dashed, (red) curved line from the bottom minterm to the top minterm helps demonstrate non-convexity when it passes through unfilled area between two filled areas.}
  \label{fig:minterm-set-operations}
\end{figure}
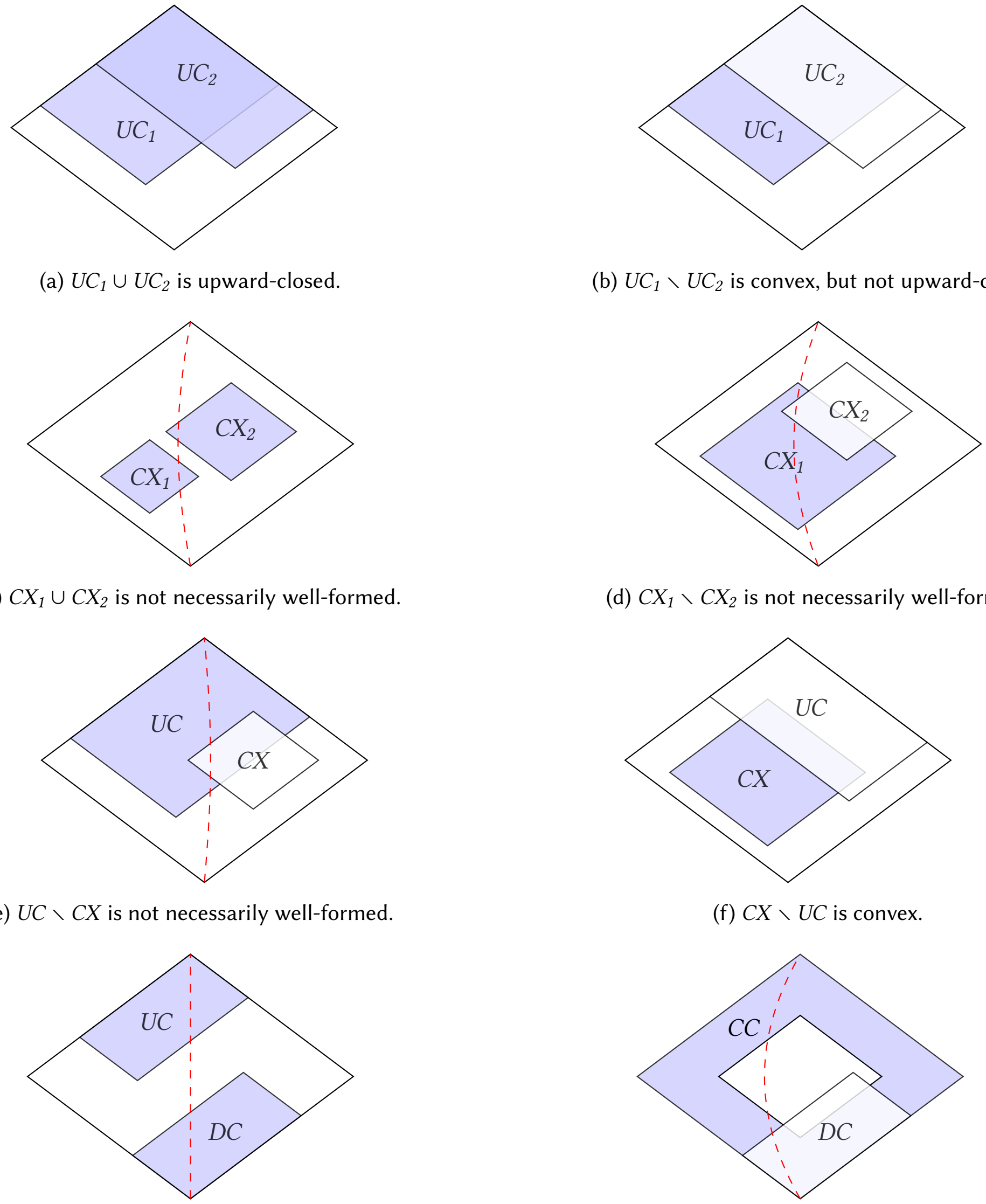


One way to inspect the operations in \Cref{tab:wellformed-operations} is to build on our illustration approach of the well-formed sets (shown earlier in \Cref{fig:minterm}).
\Cref{fig:minterm-set-operations} depicts examples corresponding to a selected subset of the results in \Cref{tab:wellformed-operations}.
A visual way to inspect the convexity of the filled area is to consider imaginary convex curves from the bottom minterm to the top minterm.
If one such curve exists that passes through the border of the filled area into the unfilled area and again into the filled area, then the filled area is not convex.
For example, that is the case in \Cref{fig:minterm-union-cx-cx} considering the union of two convex minterm sets.
Note that the complement of the filled area follows a similar pattern.
Therefore, the result is not necessarily well-formed.
\Cref{fig:minterm-union-uc-dc} shows that the union of an upward-closed set and a downward-closed set is not necessarily convex based on the red (dashed) curve.
However, its complement is convex, i.e., the result is co-convex.

Combined with the well-formed semantics of policy rules (\Cref{lem:wellformed-rules}), the above operations result in interesting well-formedness properties for rule-based policy semantics.
We refer to the well-formed structure of a rule-based policy as its shape. 
This notion of \emph{policy shape} helps with characterizing the semantics of policies and policy models.
As shown in the following theorem, different rule-based policy models result in different policy shapes.
\begin{theorem}[Policy Shapes] \label{thm:shape}
  Policy semantics expressed by certain rule-based policy models are well-formed:
  \begin{thmenum}
  \item \label{thm:shape.ddpo}
    DDPO policies are upward-closed.
  \item \label{thm:shape.dpdo}
    DPDO policies are downward-closed.
  \item \label{thm:shape.dddo}
    DDDO policies are convex.
  \item \label{thm:shape.dppo}
    DPPO policies are co-convex.
  \item \label{thm:shape.ddfa}
    DDFA policies are not necessarily well-formed.
  \item \label{thm:shape.negation}
    Negation policies are not necessarily well-formed.
  \end{thmenum} 
\end{theorem}

\begin{proof}[Proof of \Cref{thm:shape.ddpo}]
  DDPO rules have \uncomplementedexp expressions, which have upward-closed expression-minterms~(\Cref{lem:wellformed-rules}).
  The semantics of DDPO is derived based on the union of the expression-minterms of its \permit rules~(\Cref{tab:model-semantics}), which results in an upward-closed semantics (\Cref{tab:wellformed-operations.union}).
\end{proof}

\begin{proof}[Proof of \Cref{thm:shape.dpdo}]
  DPDO rules have \uncomplementedexp expressions, which have upward-closed expression-minterms~(\Cref{lem:wellformed-rules}).
  The semantics of DPDO~(\Cref{tab:model-semantics}) is derived by first calculating the union of the expression-minterms of its \deny rules, which results in an upward-closed semantics~(\Cref{tab:wellformed-operations.union}); and then subtracting that from \allminterms, which results in a downward-closed semantics~(\Cref{tab:wellformed-operations.subtraction}).
\end{proof}

\begin{proof}[Proof of \Cref{thm:shape.dddo}]
  DDDO rules have \uncomplementedexp expressions, which have upward-closed expression-minterms~(\Cref{lem:wellformed-rules}).
  The semantics of DDDO~(\Cref{tab:model-semantics}) is derived by first calculating the union of the expression-minterms of its \permit rules and \deny rules separately, which results in two upward-closed semantics (\Cref{tab:wellformed-operations.union}); and then subtracting the \deny-union semantics from the \permit-union semantics, which results in a convex semantics~(\Cref{tab:wellformed-operations.subtraction}).
\end{proof}

\begin{proof}[Proof of \Cref{thm:shape.dppo}]
  DPPO rules have \uncomplementedexp expressions, which have upward-closed expression-minterms~(\Cref{lem:wellformed-rules}).
  The semantics of DPPO~(\Cref{tab:model-semantics}) is derived by first calculating the union of the expression-minterms of its \permit rules and \deny rules separately, which results in two upward-closed semantics (\Cref{tab:wellformed-operations.union}); second, subtracting the \deny-union semantics from \allminterms, which results in a downward-closed semantics~(\Cref{tab:wellformed-operations.subtraction}); and finally, calculating the union of the previous result and \permit-union semantics calculated earlier, which results in a co-convex semantics~(\Cref{tab:wellformed-operations.union}).
\end{proof}

\begin{proof}[Proof of \Cref{thm:shape.ddfa}]
  DDFA rules have \uncomplementedexp expressions, which have upward-closed expression-minterms~(\Cref{lem:wellformed-rules}).
  The semantics of DDFA~(\Cref{tab:model-semantics}) is derived by starting from an empty set of minterms and performing a series of union and subtraction operations of upward-closed expression-minterms (corresponding to \permit and \deny rules, respectively).
  The order of those operations depends on the order of the rules in the policy.
  It is possible that the result is not necessarily well-formed.
  In particular, processing the first subtraction operation between two upward-closed semantics results in a convex semantics~(\Cref{tab:wellformed-operations.subtraction}).
  Processing any future \permit rule requires calculating the union of this convex semantics and the upward-closed semantics corresponding to the rule, which will result in a semantics that is not necessarily well-formed~(\Cref{tab:wellformed-operations.union}).
\end{proof}

\begin{proof}[Proof of \Cref{thm:shape.negation}]
  Negation rules have \compuncompexp expressions, which have convex expression-minterms~(\Cref{lem:wellformed-rules}).
  The semantics of Negation is derived based on the union of the expression-minterms of its rules~(\Cref{tab:model-semantics}), which results in a semantics that is not necessarily well-formed~(\Cref{tab:wellformed-operations.union}).
\end{proof}

\section{Expressive Power of Rule-Based Policy Models}
\label{sec:expressiveness}
The expressive power of an access control model refers to its ability to specify a diverse range of policies.
In \Cref{sec:wellformedness}, we characterized the well-formedness of the semantics that can be expressed using various rule-based policy models.
However, the characterization of the expressive power of a model is incomplete without determining what policies can (or cannot) be expressed in the context of that model.
For example, we showed that DDDO policies have convex semantics.
But, can any convex semantics be expressed using the DDDO model?
Answering such questions allows for assessing the suitability of an access control model for expressing the policies needed for a given application environment.
The following theorem establishes a set of formal results about the expressive power of the rule-based models discussed in this paper.
\begin{theorem}[Model Expressiveness] \label{thm:expressiveness}
  Rule-based models have the following expressive powers:
  \begin{thmenum}
  \item \label{thm:expressiveness.negation}
    Negation can express any semantics.
  \item \label{thm:expressiveness.ddfa}
    DDFA can express any semantics.
  \item \label{thm:expressiveness.dddo}
    DDDO can express any and only convex semantics.
  \item \label{thm:expressiveness.dppo}
    DPPO can express any and only co-convex semantics.
  \item \label{thm:expressiveness.ddpo}
    DDPO can express any and only upward-closed semantics.
  \item \label{thm:expressiveness.dpdo}
    DPDO can express any and only downward-closed semantics.
  \end{thmenum}  
\end{theorem}

We prove the correctness of each of the above results separately.
We first prove that the Negation and DDFA are the most expressive models, not only among the considered models in this paper but in comparison to any rule-based model in the context of our framework presented in \Cref{sec:models}.
\begin{proof}[Proof of \Cref{thm:expressiveness.negation}]
  Recall that Negation supports \permit rules with \compuncompexp expressions.
  Also, a condition minterm is essentially an \compuncompexp expression with all conditions present.
  Therefore, given any set of minterms as the target semantics, it suffices to specify one rule corresponding to each minterm.
  The semantics of the policy, as shown in \Cref{tab:model-semantics}, will exactly express the union of those minterms.
\end{proof}

\begin{proof}[Proof of \Cref{thm:expressiveness.ddfa}]
We propose a naive algorithm that can produce the corresponding DDFA policy to any given semantics (set of minterms).
Algorithm~\ref{alg:minterms-ddfa} loops through all possible minterms in their reverse topological order.
That means if $m_i < m_j$, $m_j$ will be considered first.
If the minterm at hand is part of the desired minterm set~$M$, a corresponding \permit rule is added.
Otherwise, a corresponding \deny rule is added.
\begin{algorithm}
  \caption{Representing a Set of Minterms using DDFA}
  \label{alg:minterms-ddfa}
  \renewcommand{\algorithmicrequire}{\textbf{Input:}}
  \begin{algorithmic}[1]
    \Require{Set of minterms $M$}
    \State $p \gets \tuple{}$
    \ForAll{$m \in \allminterms$ in reverse topological order}
      \State $\phi \gets$ conjunction of uncomplemented conditions in $m$
      \If{$m \in M$}
        \State Append \tuple{\phi, \permit{}} as last rule in $p$
        \label{alg:minterms-ddfa.permit}
      \Else
        \State Append \tuple{\phi, \deny{}} as last rule in $p$
      \EndIf
    \EndFor
    \State \Return{$p$}
  \end{algorithmic}
\end{algorithm}

\Cref{alg:minterms-ddfa} is correct if and only if $\mu(p) = M$.
We first show that given $m \in M$, $m \in \mu(p)$.
In the loop iteration corresponding to $m$, clearly $m \in \mu(\phi)$, and a corresponding rule $r=\tuple{\phi, \permit{}}$ is added to the policy.
Now, observe the DDFA semantics shown in the pseudocode in Table~\ref{tab:model-semantics}.
Rule $r$ will clearly contribute $m$ to $\mu(p)$.
We will have $m \not\in \mu(p)$ at the end of processing policy semantics only if a later processed rule $r'=\tuple{\phi', \deny{}}$ removes $m$.
In that case:
\begin{equation}
  \label{prf:ddfa-exp.1}
  m \in \mu(\phi')
\end{equation}
Let $m'$ be the denied minterm for which $r'$ was added.
Since the loop in Table~\ref{tab:model-semantics} processes rules in reverse order,
\Cref{alg:minterms-ddfa} must have appended $r'$ earlier than $r$.
Therefore, we must have $m'~\not<~m$ due to the reverse topological order.
Let $S$ and $S'$ be the set of uncomplemented conditions in $m$ and $m'$, respectively.
Since $m'~\not<~m$, we have:
\begin{equation}
  \label{prf:ddfa-exp.2}
  S' \not\subset S
\end{equation}
\Cref{prf:ddfa-exp.2} contradicts \Cref{prf:ddfa-exp.1}
which itself requires $S' \subseteq m$ by definition.
Therefore, \deny{} rule $r'$ cannot exist, and hence, $m \in \mu(p)$.
We can similarly show that if $m \in \mu(p)$ then $m \in M$.
\end{proof}

To facilitate proving the expressive power of the rest of the models, we will first show the expressive power of rule expressions in the context of those models.
\begin{lemma} \label{lem:upward-semantics-using-ncomplemented}
  Any upward-closed semantics can be expressed using the union of a set of expression-minterms of rules with \uncomplementedexp expressions.
\end{lemma}
\begin{proof}
  Let $M$ be an upward-closed set of minterms.
  Let $M_{\text{min}}$ be the minimal minterms in $M$ (i.e., $M_{\text{min}} = \set{m_i \in M \,|\, \nexists m_j \in M, m_j < m_i})$.
  Let $e_i$ be an \uncomplementedexp expression composed of uncomplemented conditions in $m_i$.
  The union of expression-minterms of such $e_i$ rules will be equivalent to $M$.
\end{proof}

We now show that DDDO and DPPO models can express convex and co-convex semantics, respectively.
\begin{proof}[Proof of \Cref{thm:expressiveness.dddo}]
  Consider a given convex set of minterms, $X$.
  We first show that $X = \upwardclosure{X} \smallsetminus Y$, where $Y$ is defined as follows:
  \begin{equation*}
    Y = \set{y \,|\, \forall x \in X: y \not\le x}
  \end{equation*}
  Suppose $x \in X$. 
  Then, clearly, $x \in \upwardclosure{X}$, and $x \not\in Y$ (based on $Y$'s definition).
  On the other hand, suppose $z \in \upwardclosure{X} \smallsetminus Y$ but $z \not\in X$ (contradiction assumption).
  Then, $z \in \upwardclosure{X} \smallsetminus X$.
  This means that $\forall x \in X, z \not\le x$,
  which subsequently means $z \in Y$.
  This is a contradiction since if $z \in Y$, then $z \not\in \upwardclosure{X} \smallsetminus Y$.
  Therefore, $X = \upwardclosure{X} \smallsetminus Y$.
  
  We note that $Y$ is upward-closed since by definition there does not exist any minterm higher than its elements.
  Recall that DDDO supports rules with only \uncomplementedexp expressions.
  A set of such rules can express upward-closed semantics $\upwardclosure{X}$ and $Y$ (based on \Cref{lem:upward-semantics-using-ncomplemented}).
  According to the semantics of DDDO policies (\Cref{tab:model-semantics}), we can define the set of rules corresponding to $\upwardclosure{X}$ and $Y$ as \permit and \deny rules, respectively, to express the semantics of $\upwardclosure{X} \smallsetminus Y$, which will be equivalent to the semantics of the given convex set of minterms, $X$.
  Therefore, DDDO policies can express any convex semantics.
  Finally, recall that the shape of DDDO policies is convex (\Cref{thm:shape.dddo}), i.e., they can express only convex semantics.
\end{proof}

\begin{proof}[Proof of \Cref{thm:expressiveness.dppo}]
  Consider a given co-convex set of minterms, $X'$.
  Its complement, $X = \allminterms \smallsetminus X'$, will be clearly convex.
  We first show that $X' = \allminterms \smallsetminus \upwardclosure{X} \, \cup \, Y$, where $Y$ is defined as follows:
  \begin{equation*}
    Y = \upwardclosure{X} \smallsetminus X
  \end{equation*}
  The correctness of the equation becomes clear by simply replacing $Y$ in the right-hand side of the equation:
  \begin{align*}
    X' &= (\allminterms \smallsetminus \upwardclosure{X}) \;\cup \; (\upwardclosure{X} \smallsetminus X) \\
       &= \allminterms \smallsetminus X
  \end{align*}
  We note that both $\upwardclosure{X}$ and $Y$ are upward-closed.
  Also, recall that DPPO supports rules with only \uncomplementedexp expressions.
  A set of such rules can express upward-closed semantics $\upwardclosure{X}$ and $Y$ (based on \Cref{lem:upward-semantics-using-ncomplemented}).
  According to the semantics of DPPO policies (\Cref{tab:model-semantics}), we can define the set of rules corresponding to $\upwardclosure{X}$ and $Y$ as \deny and \permit rules, respectively, to express the semantics of $\allminterms \smallsetminus \upwardclosure{X} \, \cup \, Y$, which will be equivalent to the semantics of the given co-convex set of minterms, $X'$.
  Therefore, DPPO policies can express any co-convex semantics.
  Finally, recall that the shape of DPPO policies is co-convex (\Cref{thm:shape.dppo}), i.e., they can express only co-convex semantics.
\end{proof}

Finally, we show that DDPO and DPDO models can express upward-closed and downward-closed semantics, respectively.

\begin{proof}[Proof of \Cref{thm:expressiveness.ddpo}]
  The semantics of a DDPO policy is derived by the union of the expression-minterms of its \permit rules (\Cref{tab:model-semantics}), each having an \uncomplementedexp expression.
  Therefore, by direct application of \Cref{lem:upward-semantics-using-ncomplemented}, DDPO can express any upward-closed semantics.
  Moreover, recall that the shape of DDPO policies is upward-closed (\Cref{thm:shape.ddpo}), i.e., they can express only upward-closed semantics.
\end{proof}

\begin{proof}[Proof of \Cref{thm:expressiveness.dpdo}]
  The semantics of a DPDO policy is derived by complementing the union of the expression-minterms of its \deny rules (\Cref{tab:model-semantics}), each having an \uncomplementedexp expression.
  The union of the expression-minterms of the rules can express any upward-closed semantics (\Cref{lem:upward-semantics-using-ncomplemented}).
  Therefore, DPDO can express any downward-closed semantics by complementing that upward-closed semantics.
  Moreover, recall that the shape of DPDO policies is downward-closed (\Cref{thm:shape.dpdo}), i.e., they can express only downward-closed semantics.
\end{proof}

\subsection{Expressiveness Hierarchy}
\label{sec:expressiveness-hierarchy}

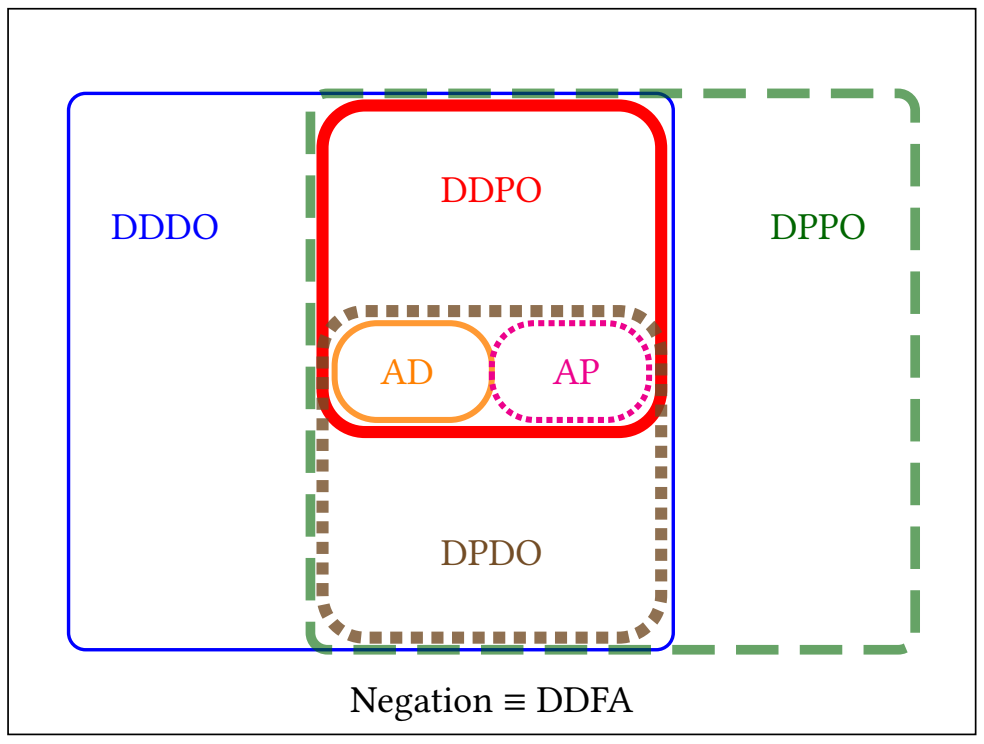
\begin{figure}
  \centering
  \scalebox{0.8}{
  \begin{tikzpicture}
    \draw (-4, -3) rectangle (4,3);
    \draw (0, -2.75) node {Negation $\equiv$ DDFA};
    \draw[blue, thick, rounded corners] (-3.5, -2.3) rectangle (1.5,2.3);
    \draw[blue] (-2.7, 1.2) node {DDDO};
    \draw[green!40!black, dash pattern=on 10pt off 5pt, line width=0.08cm, opacity=0.6, rounded corners] (3.5, -2.3) rectangle (-1.5,2.3);
    \draw[green!40!black] (2.7, 1.2) node {DPPO};
    \draw[red, line width=0.1cm, rounded corners=10pt] (-1.4, -0.5) rectangle (1.4,2.2);
    \draw[red] (0, 1.5) node {DDPO};
    \draw[brown!60!black, dotted, line width=0.1cm, opacity=0.8, rounded corners=10pt] (-1.4, -2.2) rectangle (1.4,0.5);
    \draw[brown!60!black] (0, -1.5) node {DPDO};
    \draw[orange, line width=0.05cm, opacity=0.8, rounded corners=10pt] (-1.3, -0.4) rectangle (0,0.4);
    \draw[orange] (-0.7, 0) node {AD};
    \draw[magenta, densely dotted, line width=0.05cm, rounded corners=10pt] (0, -0.4) rectangle (1.3,0.4);
    \draw[magenta] (0.7, 0) node {AP};
  \end{tikzpicture}}
  \caption{Expressiveness hierarchy of rule-based policy models (AD: all denied; AP: all permitted).}
  \label{fig:expressiveness-hierarchy}
\end{figure}

In this section, we make a few additional observations on the expressiveness results presented in \Cref{thm:expressiveness} and summarize them by building an expressiveness hierarchy.
In particular, \Cref{fig:expressiveness-hierarchy} depicts the relative expressive power of the studied rule-based models as a Venn diagram.
The proposed hierarchy not only establishes a framework for comparing the expressiveness of different models, but also provides a means to explore the expressiveness shared across them.

As shown earlier, DDPO and DPDO policies express upward-closed and downward-closed semantics, respectively.
There are only two special sets of minterms, $\allminterms$ and $\emptyset$, that are both upward-closed and downward-closed ($\upwardclosure{\allminterms} = \downwardclosure{\allminterms} = \allminterms$ and $\upwardclosure{\emptyset} = \downwardclosure{\emptyset} = \emptyset$).
Let us call their corresponding policies AP (all permitted) and AD (all denied), respectively.
Therefore, policies AP and AD are at the intersection of DDPO and DPDO policies.
Another observation is that DDPO policies (upward-closed semantics) and DPDO policies (downward-closed semantics) together form the intersection of DDDO policies (convex semantics) and DPPO policies (co-convex semantics).

\section{Policy Convertibility and Conversion}
\label{sec:conversion}
As shown in \Cref{sec:expressiveness-hierarchy}, different rule-based policy models may have overlap in their expressive power. 
In other words, the same authorization semantics might be expressible using different models.
For example, the semantics of any DDPO policy can be also expressed using semantically-equivalent DDDO or DPPO policies.
We define the \emph{policy convertibility} problem as whether a policy in a source model can be converted to a semantically-equivalent policy in a target model.
This is the decision problem corresponding to the \emph{policy conversion} problem with the goal of producing the actual policy in the target model.
Analyzing convertibility of policies is critical when considering migration to a policy model different from the current model.

In this section, we explore the policy convertibility and conversion problems in the context of the rule-based policy models studied in this paper.
The policy convertibility problem is straightforward to answer when the source model is less expressive than the target model.
The answer is always affirmative in such a case based on the expressiveness results established in this paper (e.g., from DDDO to Negation).
However, the convertibility problem becomes complicated when the source model is more expressive than the target model (e.g., from Negation to DDDO) or when the expressive power of models are not comparable (e.g., between DDDO and DPPO).
A naive approach to testing the policy convertibility (and policy conversion) is to first produce the policy minterms of the source policy, in the form of a set of condition minterms, and then test the well-formedness of that set.
However, such an approach would be computationally expensive.
The time complexity of this method will clearly be exponential in terms of the number of the conditions.
In this section, we analyze the complexity of several non-trivial convertibility problems and discuss the conversion process for the tractable cases.
\subsection{Conversion from Negation to DDDO}
\label{sec:conversion.negation-dddo}
As shown in \Cref{thm:expressiveness}, a Negation policy can express any semantics, while a DDDO policy can express any and only convex semantics.
Therefore, semantically, the \emph{Negation-to-DDDO convertibility} problem is about whether the given set of minterms implemented by a Negation policy is \emph{convex}.
A formal definition of this
decision problem is as follows:

\begin{center}
\begin{varwidth}{\columnwidth}
\noindent
\underline{\textbf{Negation-to-DDDO Convertibility}}
\begin{description}
\item[INPUT:] A Negation policy~$p$
\item[QUESTION:] Is $p$ convertible to a DDDO policy?
\end{description}
\end{varwidth}
\end{center}

Note that this problem is essentially the same as asking whether
the set of minterms defined by a DNF is convex.
We show that this is indeed a hard problem.
\begin{theorem}
  \label{thm:negation-dddo-convertibility}
  Negation-to-DDDO convertibility is co-NP-complete.
\end{theorem}
\begin{proof}
  We first show that the problem is in co-NP.
  Given a Negation policy $p_n$ that is not convertible to DDDO, there exists a certificate which consists of 3 minterms $m_1 \le m_2 \le m_3$
  where $\set{m_1, m_3} \subseteq \mu(p_n)$ and $m_2 \not\in \mu(p_n)$.
  We can verify the correctness of the certificate in polynomial time.
  The verification will take $O(|p_n|)$ by checking each given minterm against conditions of all policy rules in the worst case.
  
  We prove the co-NP-hardness of the problem by showing that the validity problem~\cite{Arora-Barak} can be polynomially reduced to Negation-to-DDDO convertibility.
  The validity problem is to determine whether a given propositional formula is evaluated to true using every assignment of truth values to its variables.
  Let $\Psi$ be a propositional formula in DNF (disjunctive normal form) over variables in set $X$, i.e., $\Psi = \Pi_1 \lor \ldots \lor \Pi_k$
  where each $\Pi_i$ is a product of variables in $X$ or their complements, e.g., $x_1\overline{x_2}x_3$ (assuming $\set{x_1, x_2, x_3} \subseteq X$).
  We define set of conditions $\mathcal{C} = \set{c_1, c_2} \cup X$,
  and form negation policy $p_n$ based on it as follows:
  \begin{policy}
  \item $\tuple{c_1c_2, \permit}$
  \item $\tuple{\overline{c_1}\overline{c_2}, \permit}$
  \item $\tuple{\Psi, \permit}$
  \end{policy}
  Note that since $\Psi$ is in DNF format, in order to conform with our framework (\Cref{sec:models}), the last rule above should be expanded into multiple rules with conditional expressions of purely conjunctive form: $\tuple{\Pi_i, \permit}$.
  However, we keep it as above for the brevity of the discussion. 
  
  We now show that $p_n$ is convertible to DDDO if and only if $\Psi$ is valid.
  Let us consider the case that $\Psi$ is valid (always equivalent to true).
  In this case, the corresponding rule in~$p_n$ always results in \permit.
  In other words,  $\mu(p_n) = \allminterms$.
  Since the policy-minterms will be convex, we can express~$p_n$ in DDDO.
  
  Now, consider the case where $\Psi$ is not valid.
  In this case, using any truth assignment, there will exist $m \in \allminterms[X]$ (set of minterms over conditions in $X$), where $m$ evaluates to false.
  Given the abovementioned first and second rule in $p_n$, we will have:
  $c_1c_2m \in \mu(p_n)$ and $\overline{c_1}\overline{c_2}m \in \mu(p_n)$.
  However, $c_1\overline{c_2}m \not\in \mu(p_n)$ since none of the rules in $p_n$ can authorize that minterm.
  Since, $\overline{c_1}\overline{c_2}m\le c_1\overline{c_2}m \le c_1c_2m$, we conclude that $\mu(p_n)$ will not be convex, and thus, it cannot be expressed in DDDO.
  This concludes our reduction and proves the co-NP-hardness of the Negation-to-DDDO convertibility problem.
\end{proof}

\subsection{Conversion from Negation to DPPO}
\label{sec:conversion.negation-dppo}
DPPO policies can express any and only co-convex semantics.
We show that \emph{Negation-to-DPPO convertibility} is also a hard problem.

\begin{theorem}
  Negation-to-DPPO convertibility is co-NP-complete.
\end{theorem}
\begin{proof}
  We first show that the problem is in co-NP.
  Given a Negation policy $p_n$ that is not convertible to DPPO, there exists a certificate which consists of 3 minterms $m_1 \le m_2 \le m_3$
  where $\set{m_1, m_3} \subseteq \allminterms \smallsetminus \mu(p_n)$ and $m_2 \in \mu(p_n)$.
  We can verify the correctness of the certificate in polynomial time by checking the minterms against policy rules.

  Similar to the proof of \Cref{thm:negation-dddo-convertibility}, we show the co-NP-hardness by reducing the validity problem~\cite{Arora-Barak} to Negation-to-DPPO convertibility.
  Let $\Psi$ be a propositional formula in DNF over variables in set $X$.
  We define set of conditions $\mathcal{C} = \set{c_1, c_2} \cup X$,
  and form negation policy $p_n$ based on it as follows:
  \begin{policy}
  \item $\tuple{\overline{c_1} c_2, \permit}$
  \item $\tuple{\Psi, \permit}$
  \end{policy}
  
  We now show that $p_n$ is convertible to DPPO if and only if $\Psi$ is valid.
  In case $\Psi$ is valid, the corresponding rule in $p_n$ always results in \permit.
  Thus, we can express $p_n$ in DPPO.
  In case $\Psi$ is not valid, using any truth assignment, there will exist minterm $m \in \allminterms[X]$, where $m$ evaluates to false.
  Consider the complement of $\mu(p_n)$.
   It will contain $\overline{c_1}\overline{c_2}m$ and $c_1c_2m$ (due to the last rule $p_n$).
  However, due to the first rule in $p_n$, the complement of $\mu(p_n)$ will not contain~$\overline{c_1}c_2m$.
  Since $\overline{c_1}\overline{c_2}m \le \overline{c_1}c_2m \le c_1c_2m$, we conclude that the complement of $\mu(p_n)$ is not convex.
\end{proof}

\subsection{Conversion from Negation to DDPO and DPDO}
\label{sec:conversion.negation-ddpo-dpdo}
Similar to \Cref{sec:conversion.negation-dddo,sec:conversion.negation-dppo}, we can show that the \emph{Negation-to-DDPO convertibility} and  \emph{Negation-to-DPDO convertibility} problems are co-NP-complete.
For brevity, we show a proof sketch of co-NP-hardness for the former problem.
Recall that DDPO can express any and only upward-closed semantics.
Let $\Psi$ be a propositional formula in DNF over variables in set $X$.
We define set of conditions $\mathcal{C} = \set{c} \cup X$, and form negation policy $p_n$ based on it as follows:
  \begin{policy}
  \item $\tuple{\overline{c}, \permit}$
  \item $\tuple{\Psi, \permit}$
  \end{policy}

If $\Psi$ is valid, then everything is permitted. 
Otherwise, there is a minterm~$m$ that does not belong to~$\mu( \Psi )$. 
In that case $\overline{c} m$ belongs to~$\mu( p_n )$ whereas $c m$ does not; thus, the semantics of~$p_n$ is not upward-closed.\\

\noindent
For the latter case, we can use the similar construction
  \begin{policy}
  \item $\tuple{{c}, \permit}$
  \item $\tuple{\Psi, \permit}$
  \end{policy}
\noindent
 over the same extended set of conditions as above.

\subsection{Conversion from DDDO to DDPO}
\label{sec:conversion.dddo-ddpo}
Recall that DDDO and DDPO policies can express all and only convex and upward-closed semantics, respectively~(\Cref{thm:expressiveness}).
Therefore, semantically, the \emph{DDDO-to-DDPO convertibility} problem is about whether a \emph{convex} set of minterms implemented using a DDDO policy is also upward-closed.
For this to be true, the semantics of the policy must either 
\begin{enumcase*}
\item include the top minterm\label{case:dddo-ddpo.topminterm}
\item be empty\label{case:dddo-ddpo.nominterm}
\end{enumcase*}.
Those cases can be easily tested for a given DDDO policy.
\Cref{case:dddo-ddpo.topminterm} can be tested by evaluating the policy for the top minterm. 
Alternatively, it can be seen that the policy must have at least one \permit rule (to permit the top minterm), and no \deny rules (to avoid overriding \permit rules on the top minterm decision).
\Cref{case:dddo-ddpo.nominterm} requires any \permit rule in the policy to be overridden by the \deny rules.
Recall that rule expressions in DDDO have upward-closed semantics (\Cref{lem:wellformed-rules}).
Whether a \deny rule overrides a \permit rule can be tested by comparing their respective conditional expressions. 
Therefore, \Cref{case:dddo-ddpo.nominterm} can be verified in polynomial time in terms of the size of the source policy and condition set \allconditions.

A conversion procedure is trivial once the convertibility is confirmed as mentioned above.
For \Cref{case:dddo-ddpo.topminterm}, the source DDDO policy can only contain \permit rules.
Therefore, the same set of rules of the source DDDO policy can be considered as the target DDPO policy.
For \Cref{case:dddo-ddpo.nominterm}, a target DDPO policy with no rules can be considered (corresponding to an empty set of minterms). 

\subsection{Conversion from DDDO to DPDO}
\label{sec:conversion.dddo-dpdo}
Recall that DDDO and DDPO policies can express all and only convex and downward-closed semantics, respectively~(\Cref{thm:expressiveness}).
Therefore, the \emph{DDDO-to-DPDO convertibility} problem is about whether the convex set of minterms implemented by a DDDO policy is also downward-closed.
Such a set must either
\begin{enumcase*}
\item include the bottom minterm\label{case:dddo-dpdo.bottomminterm}
\item be empty\label{case:dddo-dpdo.nominterm}
\end{enumcase*}.
\Cref{case:dddo-dpdo.bottomminterm} can be easily tested either by evaluating the policy for the bottom minterm, or verifying that rule \tuple{\top, \permit} exists and rule \tuple{\top, \deny} does not exist.
\Cref{case:dddo-dpdo.nominterm} will be the same as \Cref{case:dddo-ddpo.nominterm}.
Therefore, DDDO-to-DPDO convertibility can be polynomially verified.
A straightforward conversion algorithm from the source DDDO policy to the target DPDO policy follows by transferring only \deny rules in \Cref{case:dddo-dpdo.bottomminterm} and a singular rule \tuple{\top, \deny} in \Cref{case:dddo-dpdo.nominterm}.
\subsection{Conversion from DPPO to DDPO and DPDO}
\label{sec:conversion.dppo-ddpo-dpdo}
Conversion from DPPO policies (co-convex semantics) is analogous to conversion from DDDO policies (convex semantics).

With regards to the \emph{DPPO-to-DDPO convertibility} problem, the co-convex set of minterms implemented by a DPPO policy is also upward-closed if it either
\begin{enumcase*}
\item does not include the bottom minterm\label{case:dppo-ddpo.nobottomminterm}
\item includes all minterms\label{case:dppo-ddpo.allminterms}
\end{enumcase*}.
\Cref{case:dppo-ddpo.nobottomminterm} requires the existence of rule \tuple{\top, \deny} and the non-existence of rule \tuple{\top, \permit}.
\Cref{case:dppo-ddpo.allminterms} requires any \deny rule in the policy to be overridden by the \permit rules.
This can be tested by comparing the conditional expression of each \deny rule against the expressions of all \permit rules.

For the \emph{DPPO-to-DPDO convertibility} problem, the co-convex set of minterms implemented by a DPPO policy is also downward-closed if it either 
\begin{enumcase*}[resume]
\item does not include the top minterm\label{case:dppo-dpdo.notopminterm}
\item includes all minterms\label{case:dppo-dpdo.allminterms}
\end{enumcase*}.
\Cref{case:dppo-dpdo.notopminterm} requires the existence of a \deny rule and no \permit rule.
\Cref{case:dppo-dpdo.allminterms} is the same as \Cref{case:dppo-ddpo.allminterms}.

Therefore, both convertibility problems and their associated conversion have polynomial complexity.

\subsection{Conversion Between DDDO and DPPO}
\label{sec:conversion.dddo-dppo}
As shown in \Cref{sec:expressiveness-hierarchy}, neither DDDO (convex semantics) nor DPPO (co-convex semantics) is more expressive than the other.
However, the semantics that is expressible by both must be expressible by either DDPO (upward-closed semantics) or DPDO (downward-closed semantics).
In other words, a DDDO policy is convertible to a DPPO policy if and only if it is convertible to either a DDPO policy or a DPDO policy.
Therefore, \emph{DDDO-to-DPPO convertibility} can be verified by checking both DDDO-DDPO convertibility (\Cref{sec:conversion.dddo-ddpo}) and DDDO-DPDO convertibility (\Cref{sec:conversion.dddo-dpdo}).
The complexity of this task is therefore polynomial following the earlier results.
Analogously, \emph{DPPO-to-DDDO convertibility} can be verified by checking DPPO-to-DDPO convertibility and DPPO-to-DPDO convertibility.

\subsection{Summary of Convertibility Results}
\label{sec:conversion.summary}

\Cref{tab:conversion-summary} summarizes the complexity results discussed in this section.
As discussed earlier, convertibility from a less expressive model to a more expressive model is trivial.
Thus, each source model in these results (a row in the table) is either more expressive than or incomparable in terms of expressiveness to the target model (a column in the table).
We note that since DDFA has the same expressiveness power as Negation (\Cref{sec:expressiveness-hierarchy}), the convertibility results for Negation applies to DDFA as well.
The computational complexity of convertibility from the Negation and DDFA models motivates further effort to develop practical solutions, despite the inherent hardness in these cases. 

\begin{table}
  \centering
  \addtolength\tabcolsep{-2pt}
  \caption{Complexity of Convertibility to Models with Less or Incomparable Expressiveness (Convertibility from Model Indicated by Row to Model Indicated by Column).}
  \label{tab:conversion-summary}
  \begin{tabular}{c|cccc}
    & DDDO & DPPO & DDPO & DPDO \\
    \hline
    Negation/DDFA & co-NP-complete & co-NP-complete & co-NP-complete & co-NP-complete \\
    DDDO & - & P & P & P \\
    DPPO & P & - & P & P \\
  \end{tabular}
  \hspace{1.5cm}
\end{table}

\section{Related Work}
\label{sec:related}
We present the closely related literature in three areas by focusing on 1) categorizing various rule-based policy models based on factors such as complemented/uncomplemented conditions, rule modality,  default decision, and conflict resolution (as captured by our framework in Section~\ref{sec:models}); 2) policy analysis and theoretical results on the expressiveness of various access control models; and 3) efficient algorithms for migrating existing policies to a different policy model.

\subsection{Rule-Based Policy Models}\label{sec:related.models}

We studied a representative set of access control models from the literature that support our design choices discussed in Section~\ref{sec:models}.
Table~\ref{tab:policy-models} summarizes our survey results. Note that a set value for $\delta$ and $\sigma$ in the table indicates that the corresponding model could be using any one of those values given within the set (i.e., is a member of the indicated set).

\begin{table}
    \centering
    \caption{Policy Models in the Literature Mapped to our Framework Features in Terms of (Un)Complemented Conditions ($\alpha$), Rule Decision Modalities ($\beta$), Default Decision ($\delta$), And Conflict Resolution ($\sigma$). A Set Value for $\delta$ and $\sigma$ Indicates The Corresponding Model Using One Of Those Values.}
    \label{tab:policy-models}
    
    \begin{threeparttable}

    \setlength\tabcolsep{1pt} 
    \begin{tabular}{l|c|l|l|p{7cm}}
    \hline
        Work(s) & $\alpha$ & $\beta$ & $\delta$ & $\sigma$ \\
    \hline
    \hline    
        \cite{shen2006attribute} & \compuncompexp & \set{\permit} & \deny & $-$ \\
    \hline    
        \cite{ServosHGABACFormalModel2014} & \compuncompexp & \set{\permit} & \deny & $-$ \\
    \hline    
        \cite{kandala2011attribute} & \uncomplementedexp & \set{\permit, \deny} & \deny & \{\textrm{\deny-overrides}, \textrm{\permit-overrides}, \textrm{first-applicable}\} \\
    \hline    
        \cite{lang2009flexible} & \uncomplementedexp & \set{\permit, \deny} & \deny & \{\textrm{\deny-overrides}, \textrm{\permit-overrides}\} \\
    \hline    
        \cite{crampton2012ptacl} & \compuncompexp & \set{\permit, \deny} & \set{\permit, \deny} & \{\textrm{\deny-overrides}, \textrm{\permit-overrides}, \textrm{first-applicable}\} \\
    \hline    
        \cite{eXtensibleAccessControlb} & \compuncompexp & \set{\permit, \deny} & \set{\permit, \deny} & \{\textrm{\deny-overrides}, \textrm{\permit-overrides}, \textrm{first-applicable}, \textrm{only-one-applicable}\tnote{$\ddagger$}\ \,\} \\
    \hline
        \cite{masoumzadeh2010osnac} & \uncomplementedexp & \set{\permit, \deny} & \deny & \textrm{\deny-overrides} \\
    \hline
        \cite{hu2011multiparty} & \uncomplementedexp & \set{\permit, \deny} & \deny & \textrm{voting-based}\tnote{$\ddagger$} \\
    \hline    
        \cite{fong2011sacmat}\cite{bruns2012relationship} & \compuncompexp & \set{\permit} & \deny & $-$ \\
    \hline
        \cite{cheng2012relationship}\cite{cheng2014attribute} & \compuncompexp & \set{\permit} & \deny & $-$ \\
    \hline
        \cite{crampton2014path}\cite{stoller2015administrative} & \uncomplementedexp & \set{\permit, \deny} & \set{\permit, \deny} & \{\textrm{\deny-overrides}, \textrm{\permit-overrides}, \textrm{first-applicable}\} \\
    \hline
        \cite{pasarella2017datalog} & \compuncompexp & \set{\permit, \deny} & \set{\deny, \permit\tnote{$\dagger$}\ \,} & \{\textrm{\deny-overrides} , \textrm{\permit-overrides}\tnote{$\dagger$}\ \,\} \\
    \hline
        \multirow{3}{*}{\cite{bertino1999flexible}} & \uncomplementedexp & \set{\permit} & \deny & $-$ \\
         & \uncomplementedexp & \set{\deny} & \permit & $-$ \\
         & \uncomplementedexp & \set{\permit, \deny} & \deny & \{\textrm{\deny-overrides}, \textrm{most-specific-authorization}\tnote{$\ddagger$}\ \,\} \\
    \hline
        \multirow{3}{*}{\cite{jajodia1997unified}} & \compuncompexp & \set{\permit} & \deny & $-$ \\
         & \compuncompexp & \set{\deny} & \permit & $-$ \\
         & \compuncompexp & \set{\permit, \deny} & \deny & \{\textrm{\deny-overrides}, \textrm{\permit-overrides}\} \\
    \hline
    \end{tabular}
    
    \begin{tablenotes}
      \item[$\dagger$] \footnotesize{These strategies can be easily supported, yet they may not have been explicitly stated in the corresponding paper.}      \item[$\ddagger$] \footnotesize{Not currently supported in our framework as the underlying features such as indeterminate access decisions, multi-party policies, or subject hierarchies have not been captured.}
    \end{tablenotes}
    \end{threeparttable}

\end{table}

Contemporary access control models utilize the notion of authorization rules to specify which users can access what resources depending on the satisfaction of the rules' conditions. 
Two of the more recent and widely studied rule-based models are \emph{attribute-based access control} (ABAC) and \emph{relationship-based access control} (ReBAC).
ABAC policies~\cite{shen2006attribute,ServosHGABACFormalModel2014,kandala2011attribute,lang2009flexible,crampton2012ptacl,eXtensibleAccessControlb} enforce conditions on the attributes of subjects and objects to specify authorizations.
Whereas, ReBAC policies~\cite{masoumzadeh2010osnac,hu2011multiparty,fong2011sacmat,bruns2012relationship,cheng2012relationship,cheng2014attribute,crampton2014path,stoller2015administrative,pasarella2017datalog} utilize the concept of relationships, a path-based pattern of edge labels between subjects and objects, that are checked against a graph of system entities. 
Previous literature on ABAC and ReBAC models have employed either
uncomplemented conditions only (\uncomplementedexp), or both complemented and uncomplemented conditions (\compuncompexp).
In this paper, we consider both these kinds of conditional expressions during the categorization of policy models.
Besides, we take into account a special conditional expression, $\top$, that always evaluates to \emph{true}, and so is satisfied by every access request.

When a rule is applicable to an access request, the access decision associated with the rule becomes effective. We consider that a rule decision can be either \permit{} (positive authorization) or \deny{} (negative authorization). 
Several policy models allow rules with only \permit{} decisions ($\beta$=\set{\permit}) and regard everything else to be denied according to the deny-by-default strategy.
But, there are also many access control systems that support both positive and negative authorization rules ($\beta$=\set{\permit, \deny}).
In models that support both positive and negative rules, a conflict resolution strategy ($\sigma$) is used to determine the final access decision when multiple rules with different decisions are applicable to the same request.
Previous policy models have mostly employed the \deny{}-overrides, \permit{}-overrides, and first-applicable strategies to prioritize, respectively, the \deny{} decision, \permit{} decision, and decision associated with the first rule listed in a policy.
Yet, based on their underlying features, models can support additional conflict resolution techniques such as only-one-applicable (when authorization decisions can be indeterminate along with permit and deny), voting-based (to deal with authorization and privacy conflicts in a multi-party authorization setting), and most-specific-authorization-takes-precedence (when dealing with subject grouping and group hierarchy).
In this paper, in order to cover a wide range of conflict resolution strategies from the literature, we consider the \deny{}-overrides, \permit{}-overrides, and first-applicable strategies.

When no rule in a policy is applicable to an access request, a default decision ($\delta$) is employed to determine the final access decision.
Most of the policy models in the literature employ the deny-by-default strategy to adhere to the fail-safe principle. However, there are also access control systems that support the permit as well as the deny default decisions.
Besides, some models~\cite{crampton2014path,stoller2015administrative} support default decisions at the subject and object levels, in addition to the system-wide default.
In this paper, we consider that the default decision in a rule-based policy model can be set to either permit or deny (i.e., $\delta \in \set{\permit, \deny}$).

Finally, there are works~\cite{bertino1999flexible,jajodia1997unified} that, similar to our framework, can support multiple models to capture the varying protection requirements in an access control system. This, in turn, allows their frameworks to enforce different access control policies -- closed, open, and hybrid whose corresponding models are enumerated in the table -- for different classes of data objects.
Based on our expressiveness result, we note that incorporating negative authorizations (i.e., $\beta$=\set{\permit, \deny}) into access control models that already support complemented conditions (i.e., \compuncompexp) does not add to the expressiveness of the authorization model.


\subsection{Analysis of Policies and Policy Models}\label{sec:related.analysis}
We can roughly classify the literature on policy analysis into two categories: 
those analyzing policies in the context of a given access control model, and those focusing on analyzing models as a whole such as expressiveness analysis.
The former category revolves around verifying static policy properties~\cite{lin2010ijis,turkmen2017cose}, combining sophisticated policies, e.g., fine-grained integration of XACML (standard XML-based language for ABAC~\cite{eXtensibleAccessControlb}) policies in collaborative and distributed applications~\cite{rao2009sacmat}, or analyzing the expressive power of XACML's combining algorithms (conflict resolution strategy in our context)~\cite{Crampton-Williams-SACMAT}.
This policy analysis category also involves verifying the dynamic properties of an access control model that address security concerns due to prospective changes in a system's authorization state.
For example, early works focused on the well-studied \emph{safety problem}~\cite{harrison1976cacm,lipton1977linear,sandhu1988schematic,sandhu1992typed} that determines whether an access control system can reach an unsafe state in which a presumably untrusted subject has access to a certain object.
More recent works considered \emph{security analysis}~\cite{li2003sp} to verify whether security properties, other than safety, such as availability, bounded safety, and containment are maintained across state transitions in a system.
Security analysis was initially demonstrated for the RT policy languages~\cite{li2002design,li2003rt}, and later extended to other models such as RBAC~\cite{li2006tissec}, administrative RBAC~\cite{sasturkar2006csfw}, ABAC~\cite{jha2019security}, and ReBAC~\cite{masoumzadeh2018security}.

While our well-formed semantics can be used for individual policy analysis, our work is closer to the latter category which involves analyzing the expressive power of policy models.
Some works in this area have focused on expressing policies from older models in a newly proposed model to show the general applicability of the newer model.
For example, Jin et al.~\cite{jin2012dbsec} show that their attribute-based access control model, \emph{$\text{ABAC}_\alpha$}, can sufficiently express discretionary, mandatory, and role-based access control policies.
Unlike such a targeted approach, we present a general framework that allows the comparison of semantics and expressiveness among a wide range of policy models.
Our work is also distinguishable from those that target very specific expressiveness problems within a particular model.
For example, Joshi et al.~\cite{joshi2005tdsc} analyze the expressiveness of temporal constraints in the generalized temporal role-based access control model~\cite{joshi2005tkde} and show that there is a minimal set of those constraints that can sufficiently express all.

As a general framework for comparing the expressive power of policy models, Tripunitara and Li~\cite{tripunitara2007jcs} propose a theory that when one access control scheme can represent all types of policies that another can, the former is deemed to be at least as expressive as the latter.
They deal with a more dynamic version of expressiveness comparison by considering administrative policies when performing (security-preserving) simulations of one model in another (i.e., reducing the state-transition representation of policies in one scheme to that in another scheme).
A recent work utilizes their framework to show that models which support (un)complemented conditions are more expressive than models that support negative authorizations (deny rules)~\cite{iyer2022expressive}.
Compared to these works, however, we are targeting a different research question.
We are concerned with formulating expressiveness questions in terms of our unified semantics framework.
We utilize the well-formedness properties, i.e., shapes, of rule-based policies to reason about their semantics when converting a policy from one model to a semantically-equivalent policy in a different model.

\subsection{Policy Migration and Mining}\label{sec:related.migration}
Policy mining deals with migrating existing access control policies to a system that uses a different, more expressive policy model than the current one.
Policy mining is a subset of the proposed policy conversion problem.
One needs to enumerate the policy minterms of the source policy in the form of a set of condition minterms to carry out the mining task, thus undertaking a computationally expensive approach to policy conversion.
Various mining approaches have been proposed in the literature for different access control models.
In the context of role-based access control (RBAC), one aims to obtain the optimal set of roles from a given user-permission relation to decompose it into user-role and role-permission relations~\cite{LuOptimalBooleanMatrix2008,colantonio2008cost,mitra2016survey,MolloyMiningRolesNoisy2010}.
Later approaches transformed policies modeled as discretionary access control (DAC) or RBAC into attribute-based access control (ABAC) by mining rules based on attributes of users and resources~\cite{xu2014mining,iyer2018mining,alohaly2018deep,cotrini2018mining}.
Recent algorithms focused on extracting concise, high-level rules in terms of relationships between users and resources from given lower-level authorizations and entity relationships~\cite{bui2019greedy,bui2020decision,iyer2019generalized,iyer2020active}.

ReBAC policy mining is followed by the problem of determining \emph{mining feasibility,} which is a special case of our policy convertibility problem when given a relationship graph and lower-level authorizations~\cite{chakraborty2021formal,chakraborty2021feasibility}.
A recent work formulates the problem of migrating a security policy enforced in a particular device to be enforced in a different device~\cite{lobo2020security}.
They abstractly consider migrations that produce approximations of the desired policy (i.e., a stricter or a less cautious realization of the original policy) due to differences in the computational capabilities of the devices or the unavailability of information required to enforce the policy in the target device.
In contrast, in this paper, we present concrete conversion scenarios among the policy models supported in our framework, and analyze the computational complexity of convertibility when the source model has different expressiveness than the target model (e.g., from Negation to DDDO or DPPO).

In a closely related work, we previously proposed an empirical approach for Negation-to-DDDO convertibility (a co-NP-complete problem as shown in~\Cref{sec:conversion.negation-dddo}) using SAT solvers~\cite{ruiz2024sacmat}.
While that approach runs roughly in quadratic time \emph{relative to a SAT oracle,} it can still solve the problem in a reasonable time in practice.
Our current work in this paper studies the convertibility problem more generally between several other models, focusing on establishing theoretical complexity results in each case.
Exploring similar empirical solutions for other hard cases such as for Negation-to-DPPO convertibility is a promising direction for future study.


\section{Conclusion and Future Work}
\label{sec:conclusion}
We proposed a formal framework for analyzing the semantics and expressiveness of rule-based access control models and policies based on model design choices such as allowing complemented conditions, deny rules, and various conflict resolution strategies.
We also formalized the problems of policy convertibility and conversion in the context of rule-based access control  models.
We showed that rule-based semantics follow well-formedness properties that help characterize their expressiveness.
Our derived expressiveness hierarchy of rule-based models (\Cref{sec:expressiveness-hierarchy}) captures surprising results such as Negation (allowing complemented conditions, but only permit rules) having the same expressiveness as DDFA (not allowing complemented conditions, allowing both permit and deny rules, and following a first-applicable conflict resolution strategy).
Furthermore, both those models can express any rule-based semantics and, therefore, are more expressive than the other models explored in this paper.
We proved that policy convertibility is a hard problem (co-NP-complete) when converting from Negation to less-expressive models such as DDDO. 
However, convertibility can be polynomially verified (and a conversion algorithm can be proposed) for many other cases such as converting from DDDO to DPDO.
Our directions for future work include expanding policy semantics to express indeterminate condition evaluation, and exploring practical policy conversion solutions even when convertibility is a hard problem.
    




\printbibliography

@PREAMBLE{"\def\authornoop#1{}"}

@inproceedings{ruiz2024sacmat,
  title = {Converting Rule-Based Access Control Policies: From Complemented Conditions to Deny Rules},
  shorttitle = {Converting Rule-Based Access Control Policies},
  booktitle = {Proceedings of the 29th ACM Symposium on Access Control Models and Technologies},
  author = {Ruiz, Josué A. and Narendran, Paliath and Masoumzadeh, Amir and Iyer, Padmavathi},
  date = {2024-06-25},
  series = {SACMAT 2024},
  pages = {159--169},
  publisher = {Association for Computing Machinery},
  location = {New York, NY, USA},
  doi = {10.1145/3649158.3657040},
  isbn = {979-8-4007-0491-8}
}

@inproceedings{MNI21-SACMAT,
  title={Towards a Theory for Semantics and Expressiveness Analysis of Rule-Based Access Control Models},
  author={Masoumzadeh, Amirreza and Narendran, Paliath and Iyer, Padmavathi},
  booktitle={Proceedings of the 26th ACM Symposium on Access Control Models and Technologies},
  pages={33--43},
  year={2021}
}

@inproceedings{Crampton-Williams-SACMAT,
  title={On completeness in languages for attribute-based access control},
  author={Crampton, Jason and Williams, Conrad},
  booktitle={Proc. 21st ACM Symposium on Access Control Models and Technologies},
  pages={149--160},
  year={2016}
}

@book{Arora-Barak,
  title={Computational complexity: a modern approach},
  author={Arora, Sanjeev and Barak, Boaz},
  year={2009},
  publisher={Cambridge University Press}
}

@inproceedings{rao2009sacmat,
  title={An algebra for fine-grained integration of {XACML} policies},
  author={Rao, Prathima and Lin, Dan and Bertino, Elisa and Li, Ninghui and Lobo, Jorge},
  booktitle={Proc. 14th ACM Symposium on Access Control Models and Technologies},
  pages={63--72},
  year={2009}
}

@article{lin2010ijis,
  title={EXAM: a comprehensive environment for the analysis of access control policies},
  author={Lin, Dan and Rao, Prathima and Bertino, Elisa and Li, Ninghui and Lobo, Jorge},
  journal={International Journal of Information Sec.},
  volume={9},
  pages={253--273},
  year={2010},
  publisher={Springer}
}

@article{turkmen2017cose,
  title={Formal analysis of {XACML} policies using {SMT}},
  author={Turkmen, Fatih and den Hartog, Jerry and Ranise, Silvio and Zannone, Nicola},
  journal={Computers \& Security},
  volume={66},
  pages={185--203},
  year={2017},
  publisher={Elsevier}
}

@article{harrison1976cacm,
  title={Protection in operating systems},
  author={Harrison, Michael A and Ruzzo, Walter L and Ullman, Jeffrey D},
  journal={Communications of the ACM},
  volume={19},
  number={8},
  pages={461--471},
  year={1976},
  publisher={ACM New York, NY, USA}
}

@inproceedings{li2002design,
  title={Design of a role-based trust-management framework},
  author={Li, Ninghui and Mitchell, John C and Winsborough, William H},
  booktitle={Proceedings 2002 IEEE Symposium on Security and Privacy},
  pages={114--130},
  year={2002},
  organization={IEEE}
}

@inproceedings{li2003sp,
  title={Beyond proof-of-compliance: Safety and availability analysis in trust management},
  author={Li, Ninghui and Winsborough, William H},
  booktitle={2003 Symposium on Security and Privacy, 2003.},
  pages={123--139},
  year={2003},
  organization={IEEE}
}

@article{li2006tissec,
  title={Security analysis in role-based access control},
  author={Li, Ninghui and Tripunitara, Mahesh V},
  journal={ACM Transactions on Information and System Security (TISSEC)},
  volume={9},
  number={4},
  pages={391--420},
  year={2006},
  publisher={ACM New York, NY, USA}
}

@inproceedings{li2003rt,
  title={{RT}: A role-based trust-management framework},
  author={Li, Ninghui and Mitchell, John C},
  booktitle={Proceedings DARPA Information Survivability Conference and Exposition},
  volume={1},
  pages={201--212},
  year={2003},
  organization={IEEE}
}

@article{sasturkar2006csfw,
  title={Policy analysis for administrative role-based access control},
  author={Sasturkar, Amit and Yang, Ping and Stoller, Scott D and Ramakrishnan, CR},
  journal={Theoretical Computer Science},
  volume={412},
  number={44},
  pages={6208--6234},
  year={2011},
  publisher={Elsevier}
}

@inproceedings{jin2012dbsec,
  title={A unified attribute-based access control model covering {DAC, MAC and RBAC}},
  author={Jin, Xin and Krishnan, Ram and Sandhu, Ravi},
  booktitle={Data and Applications Security and Privacy XXVI: 26th Annual IFIP WG 11.3 Conference, DBSec 2012, Paris, France, July 11-13, 2012. Proceedings 26},
  pages={41--55},
  year={2012},
  organization={Springer}
}

@article{joshi2005tdsc,
  title={An analysis of expressiveness and design issues for the generalized temporal role-based access control model},
  author={Joshi, James B.D. and Bertino, Elisa and Ghafoor, Arif},
  journal={IEEE Transactions on Dependable and Secure Computing},
  volume={2},
  number={2},
  pages={157--175},
  year={2005},
  publisher={IEEE}
}

@article{joshi2005tkde,
  title={A generalized temporal role-based access control model},
  author={Joshi, James B.D. and Bertino, Elisa and Latif, Usman and Ghafoor, Arif},
  journal={IEEE Transactions on Knowledge and Data Engineering},
  volume={17},
  number={1},
  pages={4--23},
  year={2005},
  publisher={IEEE}
}

@article{tripunitara2007jcs,
  title={A theory for comparing the expressive power of access control models},
  author={Tripunitara, Mahesh V and Li, Ninghui},
  journal={Journal of Computer Security},
  volume={15},
  number={2},
  pages={231--272},
  year={2007},
  publisher={IOS Press}
}

@article{iyer2022expressive,
  title={On the Expressive Power of Negated Conditions and Negative Authorizations in Access Control Models},
  author={Iyer, Padmavathi and Masoumzadeh, Amirreza and Narendran, Paliath},
  journal={Computers \& Security},
  volume={116},
  pages={102586},
  year={2022},
  publisher={Elsevier}
}

@article{lipton1977linear,
  title={A linear time algorithm for deciding subject security},
  author={Lipton, Richard J and Snyder, Lawrence},
  journal={Journal of the ACM (JACM)},
  volume={24},
  number={3},
  pages={455--464},
  year={1977},
  publisher={ACM New York, NY, USA}
}

@article{sandhu1988schematic,
  title={The schematic protection model: its definition and analysis for acyclic attenuating schemes},
  author={{\authornoop{T}}Sandhu, Ravinderpal Singh},
  journal={Journal of the ACM (JACM)},
  volume={35},
  number={2},
  pages={404--432},
  year={1988},
  publisher={ACM New York, NY, USA}
}

@inproceedings{sandhu1992typed,
  title={The typed access matrix model},
  author={Sandhu, Ravi S},
  booktitle={Proceedings 1992 IEEE Symposium on Security and Privacy},
  pages={122--136},
  year={1992},
  organization={IEEE}
}

@article{jha2019security,
  title={Security analysis of {ABAC} under an administrative model},
  author={Jha, Sadhana and Sural, Shamik and Atluri, Vijayalakshmi and Vaidya, Jaideep},
  journal={IET information security},
  volume={13},
  number={2},
  pages={96--103},
  year={2019},
  publisher={IET}
}

@inproceedings{masoumzadeh2018security,
  title={Security analysis of relationship-based access control policies},
  author={Masoumzadeh, Amirreza},
  booktitle={Proceedings of the Eighth ACM Conference on Data and Application Security and Privacy},
  pages={186--195},
  year={2018}
}

@inproceedings{LuOptimalBooleanMatrix2008,
  title={Optimal boolean matrix decomposition: Application to role engineering},
  author={Lu, Haibing and Vaidya, Jaideep and Atluri, Vijayalakshmi},
  booktitle={IEEE 24th International Conference on Data Engineering},
  pages={297--306},
  year={2008},
  organization={IEEE}
}

@inproceedings{colantonio2008cost,
  title={A cost-driven approach to role engineering},
  author={Colantonio, Alessandro and Di Pietro, Roberto and Ocello, Alberto},
  booktitle={Proceedings of the 2008 ACM Symposium on Applied Computing},
  pages={2129--2136},
  year={2008}
}

@article{mitra2016survey,
  title={A survey of role mining},
  author={Mitra, Barsha and Sural, Shamik and Vaidya, Jaideep and Atluri, Vijayalakshmi},
  journal={ACM Computing Surveys},
  volume={48},
  number={4},
  pages={1--37},
  year={2016},
  publisher={ACM New York, NY, USA}
}

@inproceedings{MolloyMiningRolesNoisy2010,
  title={Mining roles with noisy data},
  author={Molloy, Ian and Li, Ninghui and Qi, Yuan Alan and Lobo, Jorge and Dickens, Luke},
  booktitle={Proceedings of the 15th ACM Symposium on Access Control Models and Technologies},
  pages={45--54},
  year={2010},
  organization={ACM}
}

@inproceedings{iyer2018mining,
  title={Mining positive and negative attribute-based access control policy rules},
  author={Iyer, Padmavathi and Masoumzadeh, Amirreza},
  booktitle={Proceedings of the 23nd ACM on Symposium on Access Control Models and Technologies},
  pages={161--172},
  year={2018}
}

@article{xu2014mining,
  title={Mining attribute-based access control policies},
  author={Xu, Zhongyuan and Stoller, Scott D},
  journal={IEEE Transactions on Dependable and Secure Computing},
  volume={12},
  number={5},
  pages={533--545},
  year={2014},
  publisher={IEEE}
}

@inproceedings{alohaly2018deep,
  title={A deep learning approach for extracting attributes of {ABAC} policies},
  author={Alohaly, Manar and Takabi, Hassan and Blanco, Eduardo},
  booktitle={Proceedings of the 23nd ACM on Symposium on Access Control Models and Technologies},
  pages={137--148},
  year={2018}
}

@inproceedings{cotrini2018mining,
  title={Mining {ABAC} rules from sparse logs},
  author={Cotrini, Carlos and Weghorn, Thilo and Basin, David},
  booktitle={2018 IEEE European Symposium on Security and Privacy (EuroS\&P)},
  pages={31--46},
  year={2018},
  organization={IEEE}
}

@article{bui2019greedy,
  title={Greedy and evolutionary algorithms for mining relationship-based access control policies},
  author={Bui, Thang and Stoller, Scott D and Li, Jiajie},
  journal={Computers \& Security},
  volume={80},
  pages={317--333},
  year={2019},
  publisher={Elsevier}
}

@inproceedings{bui2020decision,
  title={A Decision Tree Learning Approach for Mining Relationship-Based Access Control Policies},
  author={Bui, Thang and Stoller, Scott D},
  booktitle={Proceedings of the 25th ACM Symposium on Access Control Models and Technologies},
  pages={167--178},
  year={2020}
}

@inproceedings{iyer2019generalized,
  title={Generalized Mining of Relationship-Based Access Control Policies in Evolving Systems},
  author={Iyer, Padmavathi and Masoumzadeh, Amirreza},
  booktitle={Proc. 24th ACM Symposium on Access Control Models and Technologies},
  pages={135--140},
  year={2019},
  organization={ACM}
}

@inproceedings{iyer2020active,
  title={Active learning of relationship-based access control policies},
  author={{\authornoop{T}}Iyer, Padmavathi and Masoumzadeh, Amirreza},
  booktitle={Proc. 25th ACM Symposium on Access Control Models and Technologies},
  pages={155--166},
  year={2020}
}

@inproceedings{lobo2020security,
  title={On security policy migrations},
  author={Lobo, Jorge and Bertino, Elisa and Russos, Alessandra},
  booktitle={Proceedings of the 25th ACM Symposium on Access Control Models and Technologies},
  pages={179--188},
  year={2020}
}

@inproceedings{chakraborty2021formal,
  title={Formal analysis of {REBAC} policy mining feasibility},
  author={Chakraborty, Shuvra and Sandhu, Ravi},
  booktitle={Proc. ACM CODASPY},
  pages={197--207},
  year={2021}
}

@inproceedings{chakraborty2021feasibility,
  title={On Feasibility of Attribute-Aware Relationship-Based Access Control Policy Mining},
  author={{\authornoop{T}}Chakraborty, Shuvra and Sandhu, Ravi},
  booktitle={IFIP Annual Conference on Data and Applications Security and Privacy},
  pages={393--405},
  year={2021},
  organization={Springer}
}

@inproceedings{shen2006attribute,
  title={An attribute-based access control model for web services},
  author={Shen, Hai-bo and Hong, Fan},
  booktitle={2006 Seventh International Conference on Parallel and Distributed Computing, Applications and Technologies (PDCAT'06)},
  pages={74--79},
  year={2006},
  organization={IEEE}
}

@inproceedings{ServosHGABACFormalModel2014,
  title={{HGABAC}: Towards a formal model of hierarchical attribute-based access control},
  author={Servos, Daniel and Osborn, Sylvia L},
  booktitle={International Symposium on Foundations and Practice of Security},
  pages={187--204},
  year={2014},
  organization={Springer}
}

@inproceedings{kandala2011attribute,
  title={An attribute based framework for risk-adaptive access control models},
  author={Kandala, Savith and Sandhu, Ravi and Bhamidipati, Venkata},
  booktitle={2011 Sixth International Conference on Availability, Reliability and Security},
  pages={236--241},
  year={2011},
  organization={IEEE}
}

@article{lang2009flexible,
  title={A flexible attribute based access control method for grid computing},
  author={Lang, Bo and Foster, Ian and Siebenlist, Frank and Ananthakrishnan, Rachana and Freeman, Tim},
  journal={Journal of Grid Computing},
  volume={7},
  number={2},
  pages={169},
  year={2009},
  publisher={Springer}
}

@misc{eXtensibleAccessControlb,
  title = {{{eXtensible Access Control Markup Language}} ({{XACML}}) {{Version}} 3.0},
  howpublished = {\url{http://docs.oasis-open.org/xacml/3.0/xacml-3.0-core-spec-os-en.html}},
  year = 2013
}

@inproceedings{crampton2012ptacl,
  title={{PTaCL}: A language for attribute-based access control in open systems},
  author={Crampton, Jason and Morisset, Charles},
  booktitle={International Conference on Principles of Security and Trust},
  pages={390--409},
  year={2012},
  organization={Springer}
}

@inproceedings{masoumzadeh2010osnac,
  title={{OSNAC}: An ontology-based access control model for social networking systems},
  author={Masoumzadeh, Amirreza and Joshi, James},
  booktitle={2nd International Conference on Social Computing},
  pages={751--759},
  year={2010},
  organization={IEEE}
}

@inproceedings{hu2011multiparty,
  title={Multiparty authorization framework for data sharing in online social networks},
  author={Hu, Hongxin and Ahn, Gail-Joon},
  booktitle={IFIP Annual Conf. Data and Applications Security and Privacy},
  pages={29--43},
  year={2011},
  organization={Springer}
}

@inproceedings{fong2011sacmat,
  title={Relationship-based access control policies and their policy languages},
  author={Fong, Philip WL and Siahaan, Ida},
  booktitle={Proc. 16th ACM Symposium on Access Control Models and Technologies},
  pages={51--60},
  year={2011}
}

@inproceedings{bruns2012relationship,
  title={Relationship-based access control: its expression and enforcement through hybrid logic},
  author={Bruns, Glenn and Fong, Philip WL and Siahaan, Ida and Huth, Michael},
  booktitle={Proceedings of the Second ACM Conference on Data and Application Security and Privacy},
  pages={117--124},
  year={2012}
}

@inproceedings{cheng2012relationship,
  title={Relationship-based access control for online social networks: Beyond user-to-user relationships},
  author={Cheng, Yuan and Park, Jaehong and Sandhu, Ravi},
  booktitle={2012 International Conference on Privacy, Security, Risk and Trust},
  pages={646--655},
  year={2012},
  organization={IEEE}
}

@inproceedings{cheng2014attribute,
  title={Attribute-aware relationship-based access control for online social networks},
  author={{\authornoop{T}}Cheng, Yuan and Park, Jaehong and Sandhu, Ravi},
  booktitle={IFIP Annual Conference on Data and Applications Security and Privacy},
  pages={292--306},
  year={2014},
  organization={Springer}
}

@inproceedings{crampton2014path,
  title={Path conditions and principal matching: a new approach to access control},
  author={Crampton, Jason and Sellwood, James},
  booktitle={Proceedings of the 19th ACM Symposium on Access Control Models and Technologies},
  pages={187--198},
  year={2014}
}

@inproceedings{stoller2015administrative,
  title={An administrative model for relationship-based access control},
  author={Stoller, Scott D},
  booktitle={IFIP Annual Conference on Data and Applications Security and Privacy},
  pages={53--68},
  year={2015},
  organization={Springer}
}

@inproceedings{pasarella2017datalog,
  title={A datalog framework for modeling relationship-based access control policies},
  author={Pasarella, Edelmira and Lobo, Jorge},
  booktitle={Proceedings of the 22nd ACM on Symposium on Access Control Models and Technologies},
  pages={91--102},
  year={2017}
}

@article{bertino1999flexible,
  title={A flexible authorization mechanism for relational data management systems},
  author={Bertino, Elisa and Jajodia, Sushil and Samarati, Pierangela},
  journal={ACM Transactions on Information Systems},
  volume={17},
  number={2},
  pages={101--140},
  year={1999},
  publisher={ACM New York, NY, USA}
}

@article{jajodia1997unified,
  title={A unified framework for enforcing multiple access control policies},
  author={Jajodia, Sushil and Samarati, Pierangela and Subrahmanian, VS and Bertino, Eliza},
  journal={ACM Sigmod Record},
  volume={26},
  number={2},
  pages={474--485},
  year={1997},
  publisher={ACM New York, NY, USA}
}
\end{document}